\documentclass[fleqn,usenatbib]{mnras}

\usepackage{newtxtext,newtxmath}

\usepackage[T1]{fontenc}

\usepackage{graphicx}	
\usepackage{amsmath}	
\usepackage{newtxtext,newtxmath}

\usepackage{natbib}
\usepackage{hyperref}
\usepackage{bm}
\usepackage{xcolor}
\usepackage{xspace}

\newcommand{\planck}{{\it Planck}}

\newcommand{\xmm}{{\it XMM-Newton}}

\newcommand{\beq}{\begin{equation}}
\newcommand{\eeq}{\end{equation}}
\newcommand{\beqa}{\begin{eqnarray}}
\newcommand{\eeqa}{\end{eqnarray}}

\newcommand{\msun}{${\rm M}_{\odot}$}

\newcommand{\eg}{e.g.,\xspace}
\newcommand{\ie}{i.e.\xspace}

\def\der{{\rm d}}

\title[Constraining cosmology with a new all-sky $y$-map]{Constraining cosmology with a new all-sky Compton parameter map \\ from the \planck\ PR4 data}

\author[Tanimura et al.]{Hideki Tanimura,$^{1}$\thanks{E-mail: hideki.tanimura@ias.u-psud.fr}
Marian Douspis,$^{1}$
Nabila Aghanim,$^{1}$
and Laura Salvati$^{1,2}$
\\
$^{1}$Universit\'{e} Paris-Saclay, CNRS, Institut d'Astrophysique Spatiale, B\^atiment 121, 91405 Orsay, France\\
$^{2}$INAF – Osservatorio Astronomico di Trieste, Via G. B. Tiepolo 11, 34143 Trieste, Italy
}

\date{Accepted XXX. Received YYY; in original form ZZZ}

\pubyear{2021}

\begin{document}
\label{firstpage}
\pagerange{\pageref{firstpage}--\pageref{lastpage}}
\maketitle

\begin{abstract}
We constructed a new all-sky Compton parameter map ($y$-map) of the thermal Sunyaev-Zel'dovich (tSZ) effect from the 100 to 857 GHz frequency channel maps delivered within the \planck\ data release 4. The improvements in terms of noise and systematic effects translated into a $y$-map with a noise level smaller by $\sim$7\% compared to the maps released in 2015, and with significantly reduced survey stripes. The produced 2020 $y$-map is also characterized by residual foreground contamination, mainly due to thermal dust emission at large angular scales and to CIB and extragalactic point sources at small angular scales. Using the new \planck\ data, we computed the tSZ angular power spectrum and found that the tSZ signal dominates the $y$-map in the multipole range, 60 < $\ell$ < 600. We performed the cosmological analysis with the tSZ angular power spectrum and found $S_8 = 0.764 \, _{-0.018}^{+0.015} \, (stat) \, _{-0.016}^{+0.031} \, (sys) $, including systematic uncertainties from a hydrostatic mass bias and pressure profile model. The $S_8$ value may differ by $\pm0.016$ depending on the hydrostatic mass bias model and by $+0.021$ depending on the pressure profile model used for the analysis. The obtained value is fully consistent with recent KiDS and DES weak-lensing observations. While our result is slightly lower than the \planck\ CMB one, it is consistent with the latter within 2$\sigma$.
\end{abstract}





\section{Introduction}
\label{sec:intro}

The thermal Sunyaev-Zel'dovich (tSZ) effect \citep{Sunyaev1972} is due to the inverse Compton scattering of cosmic microwave background (CMB) photons by hot electrons along the line of sight and in particular in clusters of galaxies. The tSZ effect has been measured by the \planck\ satellite, the Atacama Cosmology Telescope (ACT), and the South Pole Telescope (SPT) in a large field, and Compton parameter maps (\ie, \citealt{Planck2014XXI, Planck2016XXII, Aghanim2019, Madhavacheril2020, Bleem2021}), referred to as $y$-map, have been constructed thanks to optimized component separation methods (\ie, \citealt{Remazeilles2011, Hurier2013, Bourdin2020, Bonjean2020}).

Especially, \planck\ produced an all-sky $y$-map using the public data release 2 (PR2) and the $y$-map was extensively used for a wide range of studies: the detection of galaxy clusters \citep{Planck2016XXVII}, baryonic physics in galaxy groups and clusters (\ie, \citealt{Greco2015, Hill2018, Tanimura2020a}), the distribution of diffuse gas in the large-scale structure (\ie, \citealt{Tanimura2019b, Graaff2019, Tanimura2019a, Tanimura2020b}), cross-correlation analyses (\ie, \citealt{Hojjati2017, Makiya2018, Chiang2020, Koukoufilippas2020, Osato2020}) 
and the constraints on cosmological parameters (\ie, \citealt{Planck2016XXIV, Hurier2017, Bolliet2018, Salvati2018}).

The current standard cosmological model, $\Lambda$ cold dark matter ($\Lambda$CDM) model mainly consisting of dark energy and dark matter, provides a wonderful fit to many cosmological data \citep{Planck2020I}. However, a slight discrepancy was found in the latest data analyses between the CMB anisotropies ($z\sim1100$) and other low-redshift ($z\sim0-1$) cosmological probes for the $S_8 (\equiv \sigma_8 (\Omega_m / 0.3)^{0.5})$ cosmological parameter:  the amplitude of matter density fluctuations, $\sigma_8$, scaled by the square root of the matter density, $\Omega_m$. 
For example, the $S_8$ value was precisely measured to be $S_8 = 0.830 \pm 0.013$ with the \planck\ CMB observation \citep{Planck2020VI}. However, a lower value of $\sigma_8 = 0.75 \pm 0.03$ was measured using the population of galaxy clusters detected by \planck\ at low redshift ($z<0.6$) \citep{Planck2014XX, Planck2016XXIV}, thus called ``$S_8$ tension''. Furthermore, other low-redshift observations with gravitational lensing by the Kilo-Degree Survey (KiDS) \citep{Heymans2021} and the Dark Energy Survey (DES) \citep{DES2018, DES2021} experiments also found lower $S_8$ values.
These results may indicate that the cosmic structure growth is slower than the prediction based on the CMB measurement 
and may require modifications to the standard cosmological model such as dark energy interaction (\ie, \citealt{Valentino2020}) and dark matter decay (\ie,  \citealt{Berezhiani2015, Pandey2020}) to explain all these measurements. 

In this paper, we reconstructed a new all-sky $y$-map using the \planck\ frequency maps from the public data release 4 (\citealt{Planck2020LVII}; PR4). The paper is structured as follows. Section \ref{sec:data} briefly describes the \planck\ data used to reconstruct the $y$-map. Section \ref{sec:mapmake} presents the reconstruction method. Then, we discuss the validation of the reconstructed $y$-map in the pixel domain, including signal and noise characterizations in Sect. \ref{sec:anapix} and the angular power spectrum in Sect. \ref{sec:anaps}. We present the cosmological interpretation of the power spectrum in Sect. \ref{sec:cosmo} and end with the conclusions in Sect. \ref{sec:summary}.


\section{\planck\ PR4 data}
\label{sec:data}

The \planck\ PR4 data implemented several improvements with respect to the previous data release: the usage of foreground polarization priors during the calibration stage to break scanning-induced degeneracies, the correction of bandpass mismatch at all frequencies, and the inclusion of 8\% of data collected during repointing maneuvers, etc.
These improvements reduced both noise and systematic effects in the frequency maps at essentially all angular scales and yielded better internal consistency between the various frequency bands. 
The \planck\ PR4 data also performed the first estimate of the Solar dipole determined through component separation across all nine \planck\ frequencies, which was removed from the original band maps before the $y$-map reconstruction process. 

The band maps are provided in HEALPix format \citep{Gorski2005} at Nside = 2048. The FWHM of the beam and conversion factors of the Compton $y$ parameter at the \planck\ bands are described in Table 1 in \cite{Planck2016XXII}. 
\planck\ team also split data into two for noise characterization and produced two maps at each frequency, called half-ring maps. The difference between the two half-ring maps at a given frequency can be used as noise maps of the band maps.
The astrophysical emissions are canceled out in the noise map, representing a statistical instrumental noise. We used the noise maps in the $y$-map reconstruction process. 


\section{Reconstruction of all-sky tSZ map}
\label{sec:mapmake}

\subsection{Compton $y$ parameter}

The Compton $y$ parameter is proportional to the line-of-sight integral of electron pressure, $P_{\rm e}=n_{\rm e}k_{\rm B}T_{\rm e}$, where $n_{\rm e}$ is the physical electron number density,  $k_{\rm B}$ is the Boltzmann constant, and $T_{\rm e}$ is the electron temperature. In an angular direction of $\bm{\hat{n}}$, it is expressed by 
\beq
    y(\bm{\hat{n}}) = \frac{\sigma_{\rm T} }{m_{\rm e} {\it c}^2} 
    \int P_{\rm e}(\bm{\hat{n}}) \, \der l\ ,
\label{eq1}
\eeq
where $\sigma_{\rm T}$ is the Thomson cross section, $m_{\rm e}$ is the mass of electron, $c$ is the speed of light, and $l$ is the {\it physical} distance. 
The change to the CMB temperature by the tSZ effect, $\Delta T$, at frequency $\nu$ is given by 
\beq
    \frac{\Delta T}{T_{\rm CMB}}(\nu,\bm{\hat{n}}) = f(x) \, y(\bm{\hat{n}}), 
\label{eq2}
\eeq
where $T_{\rm CMB}$ is the CMB temperature.
The frequency dependence of the tSZ effect is included in the pre-factor $f(x)$ as 
\beq
    f(x) = x \ \mathrm{coth} \left(\frac{x}{2} \right) - 4 \quad \left(x = \frac{h \nu}{k_{\rm B} T_{\rm CMB}} \right)
\label{eq3}
\eeq
in the thermodynamic temperature unit, where $h$ is the Planck constant. We ignore relativistic corrections to the tSZ spectrum.

\subsection{tSZ reconstruction}
\label{subsec:reconstruction}

The tSZ signal is subdominant relative to the CMB and other foreground emissions in the \planck\ frequency bands. Thus tailored component separation algorithms are required to reconstruct the tSZ map (\ie, \citealt{Remazeilles2011, Hurier2013, Bourdin2020, Bonjean2020}). We adopted the MILCA (Modified Internal Linear Combination Algorithm) \citep{Hurier2013} used for one of the \planck\ $y$-map reconstructions in \cite{Planck2016XXII} and mostly followed their reconstruction procedure (the difference in the procedure is summarized at the end of Sect. \ref{subsec:reconstruction}). The MILCA is based on the Internal Linear Combination (ILC) approach that preserves an astrophysical component given the known spectrum by minimizing the variance in the reconstructed signal (here, the tSZ effect). Additional constraints were included: removal of the CMB with an extra one degree of freedom and minimization of noises with extra two degrees of freedom. For the noise minimization, the noise maps in Section \ref{sec:data} were used for the evaluation. 

Six HFI maps from 100 to 857 GHz were used and convolved to a common resolution of 10'.
The usage of the dust-dominant 857 GHz map was restricted only at multipoles $\ell$ < 300 to minimize residuals from Infrared (IR) point sources and Cosmic Infrared Background (CIB) emission in the final $y$-map.

\begin{figure}
    \centering
    \includegraphics[width=\linewidth]{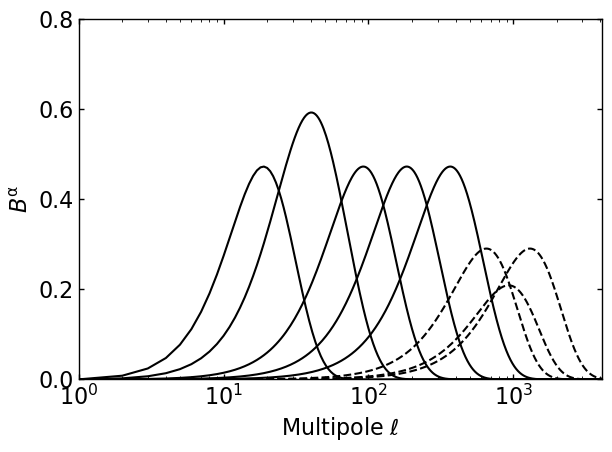}
    \caption{Multipole window functions used with the MILCA algorithm. The 857 GHz map is not used at small scales corresponding to the window functions in dashed lines. }
    \label{fig:scale-filter}
    \end{figure}

ILC weights were allowed to vary as a function of multipole $\ell$ to incorporate the scale-dependence of the reconstructed signal. Nine filters in $\ell$ space, as shown in Fig.~\ref{fig:scale-filter}, were used. These filters have an overall transmission of one, except for $\ell$ < 60. These large angular scales are dominated by foreground emissions. Therefore we optimized a Gaussian filter to reduce their residuals. 
ILC weights were also allowed to vary in different sky regions. Thus, ILC weights were computed independently in different sky regions in different multipole maps. The spatial segmentation was optimized for each multipole map with a minimum of 12 regions at low multipoles and 3072 regions at high multipoles.

In addition to the signal $y$-map, we also produced the so-called first and last (F and L hereafter) $y$-maps from the half-ring maps. We used these maps for the noise estimate in the final $y$-map in Sect. \ref{sec:anapix} and for the power spectrum estimation in Sect. \ref{sec:anaps}. 
Moreover, we produced, in the same manner, a $y$-map with a higher angular resolution of 7.5'. Note that this map is discussed in Appendix. \ref{sec:ymap7p5} and we mainly discuss the 10' $y$-map in the rest of this paper.

In summary, we followed the MILCA algorithm in \cite{Hurier2013} and reconstructed 10' and 7.5' resolution $y$-maps with Nside = 2048. 
There are two main differences compared to \cite{Planck2016XXII} in our process: We used the public multipole filters used for NILC in \cite{Planck2016XXII}, and we optimized a Gaussian filter to reduce foreground residuals, in which an overall transmission was one at $\ell$ > 8 in \cite{Planck2016XXII}, but one at $\ell$ > 60 in this work.

\subsection{Reconstructed tSZ map}

\begin{figure*}
    \centering
    \includegraphics[width=0.8\linewidth]{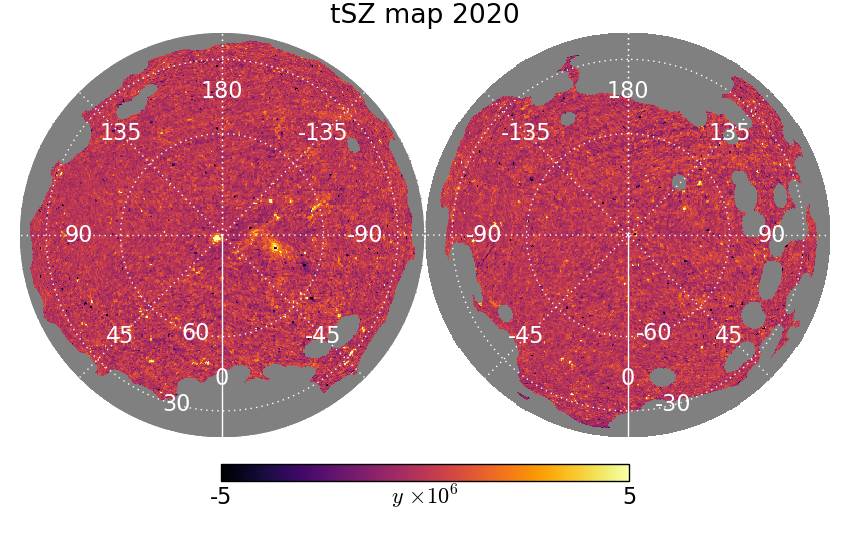}
    \includegraphics[width=0.8\linewidth]{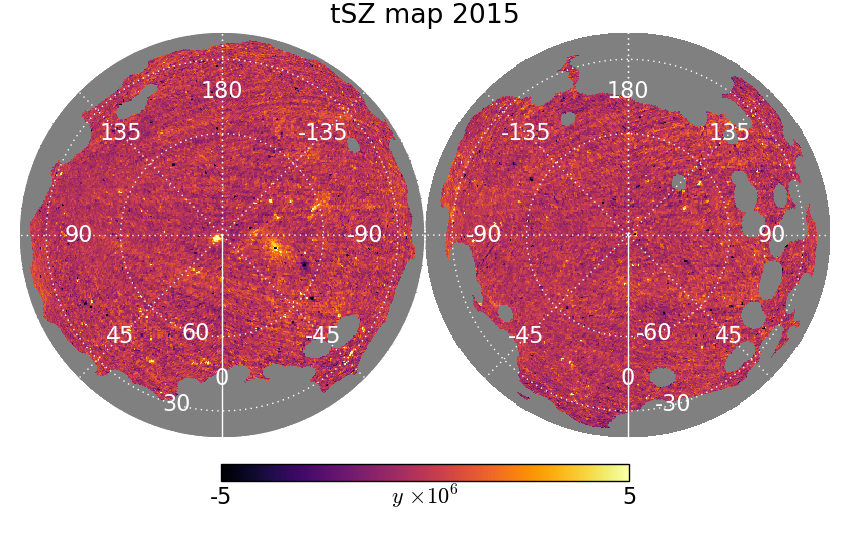}
    \caption{All-sky Compton $y$ parameter maps reconstructed in this paper ({\it top}) with the MILCA algorithm and in \protect\cite{Planck2016XXII} ({\it bottom}) in orthographic projections. The North Galactic hemisphere centered at the pole is shown on the left, and the South is on the right. Longitude=0 is shown in white solid lines and other longitudes and latitudes are shown in white dashed lines. The pixel resolution is chosen to be Nside = 128 for visualization purposes. Compact bright spots in the maps are clusters of galaxies observed via the tSZ effect, whereas compact dark spots are radio sources. }
    \label{fig:ymap}
    \end{figure*}

Figure.~\ref{fig:ymap} shows the all-sky Compton $y$ parameter map reconstructed in this paper (hereafter referred to as {\it 2020 $y$-map}; top panel) and in \cite{Planck2016XXII} with the MILCA algorithm (hereafter referred to as {\it 2015 $y$-map}; bottom panel).  The North Galactic hemisphere centered at the pole is shown on the left, and the South is on the right.
The same 40\% of the sky region around the Galactic plane is masked in both maps displayed with pixel resolution Nside = 128 for visualization purposes.

Clusters of galaxies appear as positive sources, and the Coma cluster and Virgo supercluster are visible near the north Galactic pole. Astrophysical contaminations such as strong Galactic and extragalactic radio sources show up as negative bright spots and are masked before any scientific analysis in this paper. Residual Galactic contamination is visible as diffuse positive structures around the edges of the masked area in both maps. Residuals from systematic effects along the scanning direction show up as stripes. The stripe level is significantly reduced in the 2020 $y$-map. The reduction is a direct consequence of the more efficient de-striping procedure of the scanning pattern in the original band maps released by \cite{Planck2020LVII}. 

\section{Pixel space analysis}
\label{sec:anapix}

\subsection{Signal distribution on the $y$-map}
\label{subsec:ysignal}

\begin{figure}
    \centering
    \includegraphics[width=\linewidth]{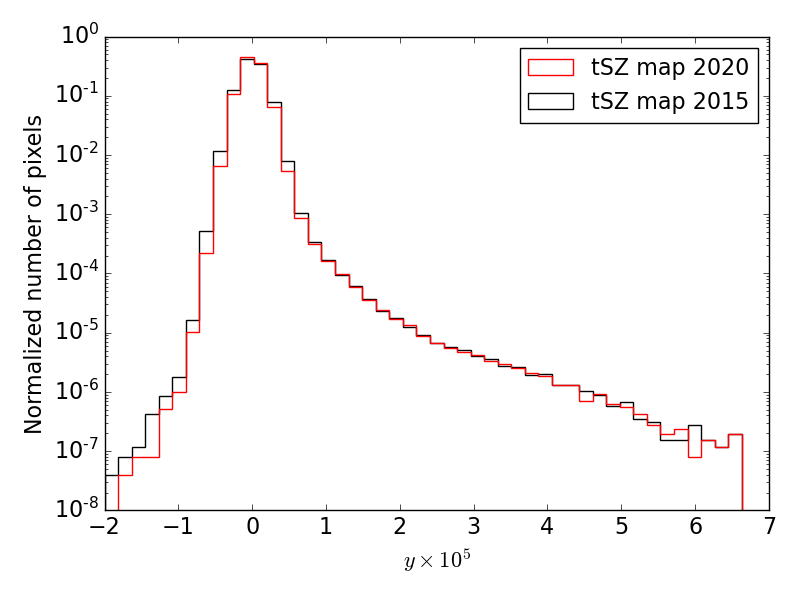}
    \caption{Histograms of the 2020 and 2015 $y$-maps. The y-axis is shown on a logarithmic scale. }
    \label{fig:hist_ysignal}
    \end{figure}

The $y$-map pixel distribution is expected to have an asymmetric behavior with a significant positive tail associated with the tSZ signal from clusters of galaxies and hot gas in the large-scale structure of the Universe \citep{Tanimura2020b, Tanimura2020c}. 
Fig.~\ref{fig:hist_ysignal} shows this in the histogram of the $y$-map after masking the Galactic plane contribution. The $y$-axis is shown on a logarithmic scale to clarify the positive tail. 
We compared the histogram with the one from the 2015 $y$-map. 
The positive tails are consistent between the 2020 and 2015 $y$-maps, suggesting that the extracted tSZ signals are consistent with each other.
On the other hand, the amplitude of the negative tail is lower in the 2020 $y$-map, which is possibly due to lower noise and/or lower radio source contamination. 

\subsection{tSZ signal from resolved sources}

\begin{figure}
    \centering
    \includegraphics[width=\linewidth]{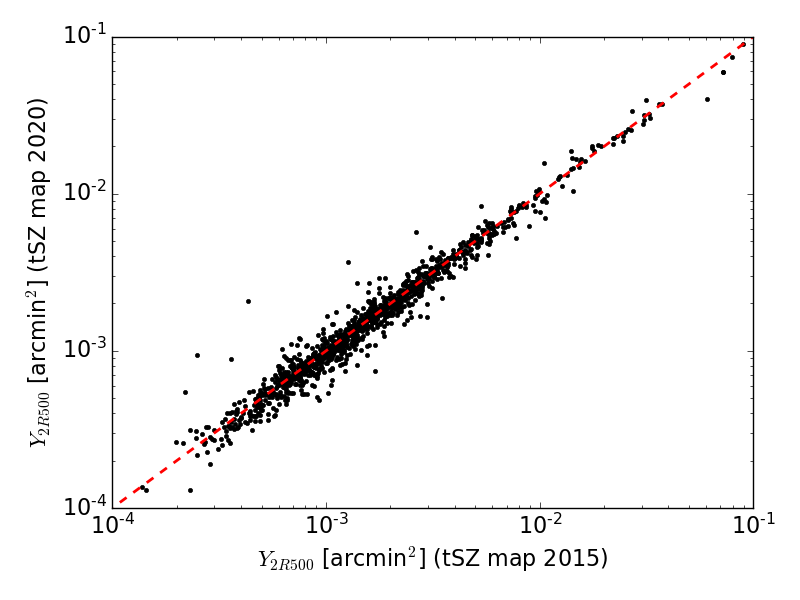}
    \caption{Comparison of the tSZ fluxes of 957 PSZ2 clusters within $2 \times R_{500}$ ($Y_{2R_{500}}$) estimated on the 2020 and 2015 $y$-maps with aperture photometry. Red dashed line is the case when $Y_{2R_{500}}$ values from the 2020 and 2015 $y$-maps are equal.}
    \label{fig:y2r500}
    \end{figure}

The \planck\ team detected 1,653 tSZ sources based on the matched filtering technique \citep{Haehnelt1996} applied for the \planck\ PR2 band maps (\citealt{Planck2016XXVII}; PSZ2 clusters).
We compared the tSZ fluxes of the PSZ2 clusters on the 2020 and 2015 $y$-maps. The tSZ fluxes were computed with aperture photometry by summing the pixel values within $2 \times R_{500}$ ($Y_{2R_{500}}$) on the $y$-maps, corresponding to about their virial radius. Note that only 957 out of 1,653 clusters are used in the comparison because others do not have mass or radius estimates. The result is shown in Fig.~\ref{fig:y2r500}. The values from the 2020 $y$-map are in good agreement with the ones from the 2015 $y$-map, with the slope of $0.995 \pm 0.006$ and slightly lower intercept by a factor of 1.9\% in a linear fit. The slight shift to lower values is possibly due to the lower noise and reduced stripes in the new $y$-map.

\begin{figure*}
    \centering
    \includegraphics[width=\linewidth]{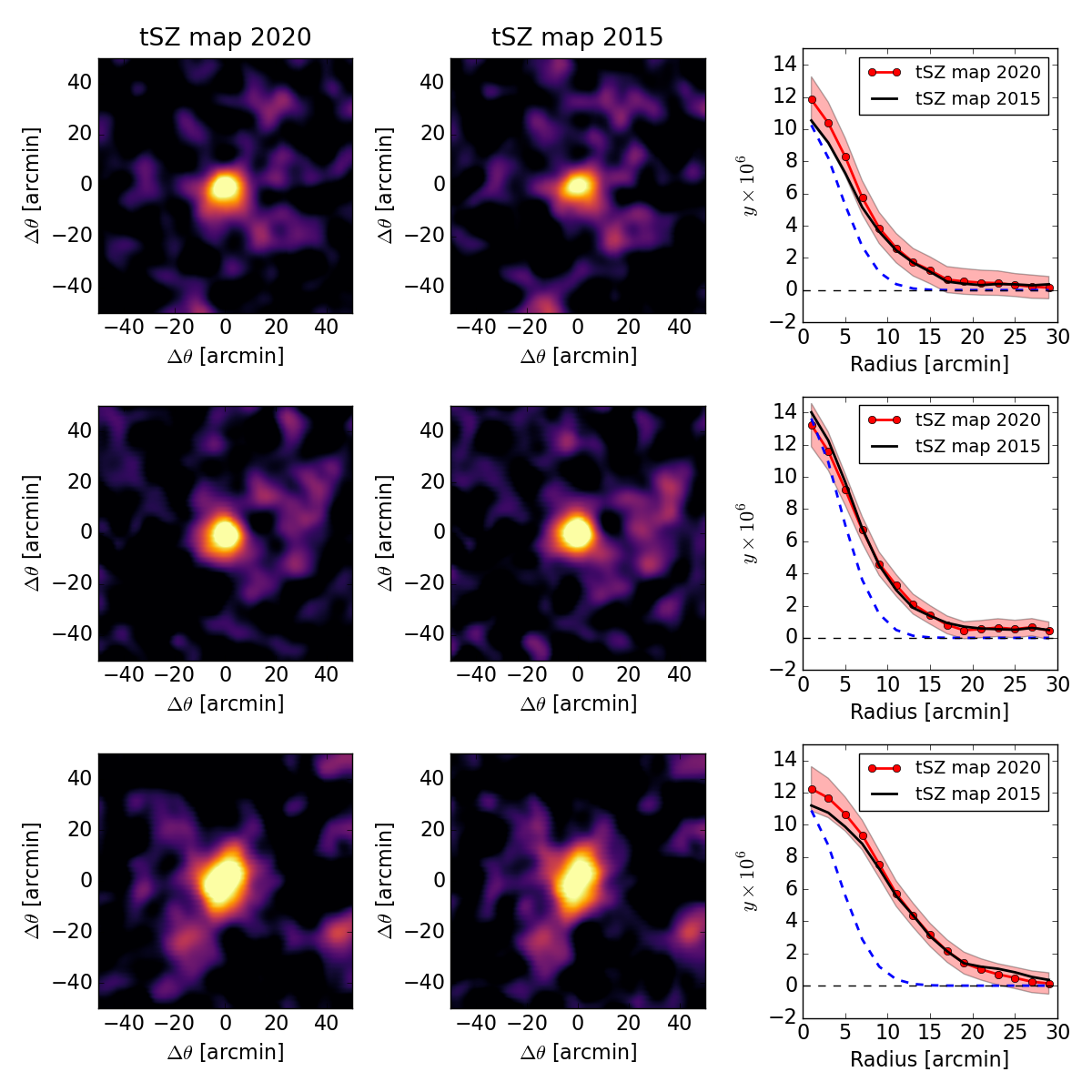}
    \caption{Two-dimensional $y$-maps centered at the position of three PSZ2 clusters corresponding to Abell 2397 at $z=0.22$ (Top row), RXC J2104.3-4120 at $z=0.16$ (middle row) and Abell 2593 at $z=0.04$ (bottom row) extracted from the 2020 (left column) and 2015 (middle column) $y$-maps. Radial profiles of these clusters obtained from the 2020 (red lines) and 2015 (black lines) $y$-maps with 10' Gaussian beams in blue dashed lines (right column). The 1$\sigma$ uncertainties of the radial profiles from the 2020 $y$-map are shown in a red shaded area. }
    \label{fig:yprof}
    \end{figure*}

We highlight in more detail some PSZ2 clusters. 
Figure.~\ref{fig:yprof} presents two-dimensional $y$-maps centered at the position of three PSZ2 clusters corresponding to Abell 2397 at $z=0.22$ (top row), RXC J2104.3-4120 at $z=0.16$ (middle row), and Abell 2593 at $z=0.04$ (bottom row), using the 2020 (left column) and 2015 (middle column) $y$-maps. In the right column of the figure, we compare the radial profiles of these clusters obtained from the 2020 (red lines) and 2015 (black) $y$-maps. The 1$\sigma$ uncertainties of the radial profiles from the 2020 $y$-map are shown in a red shaded area. For reference, we also show the radial profile of a 10' Gaussian beam in blue dashed lines. 
We found that the images are visually similar and the profiles are consistent within their 1$\sigma$ uncertainties using these two $y$-maps.

\begin{figure*}
    \centering
    \includegraphics[width=\linewidth]{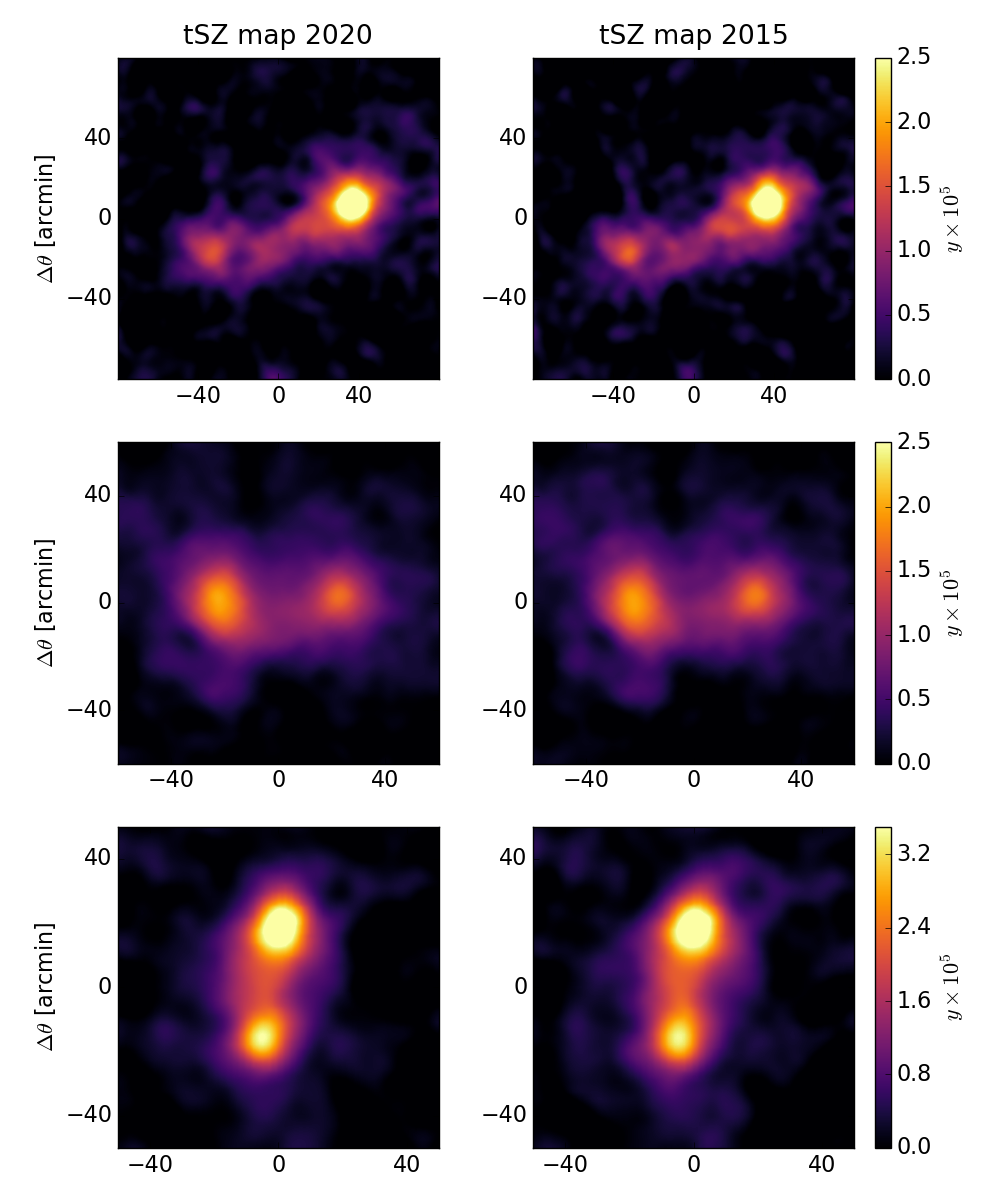}
    \caption{Two-dimensional $y$-maps centered at the position of three cluster systems: Shapley supercluster (top row), A3395-A3391 system (middle row) and A399-A401 cluster pair (bottom row) extracted from the 2020 (left) and 2015 (right) $y$-maps. }
    \label{fig:yfil}
    \end{figure*}

We also checked tSZ signals from fainter emission in three well-known cluster systems: Shapley supercluster, A3395-A3391 system, and the cluster pair A399-A401. Figure.~\ref{fig:yfil} presents two-dimensional $y$-maps around these three systems using 2020 (left) and 2015 (right) $y$-maps. The tSZ signals in the bridge regions are visible in both $y$-maps.

\subsection{Noise distribution on the $y$-map}
\label{subsec:ynoise}

\begin{figure*}
    \centering
    \includegraphics[width=0.49\linewidth]{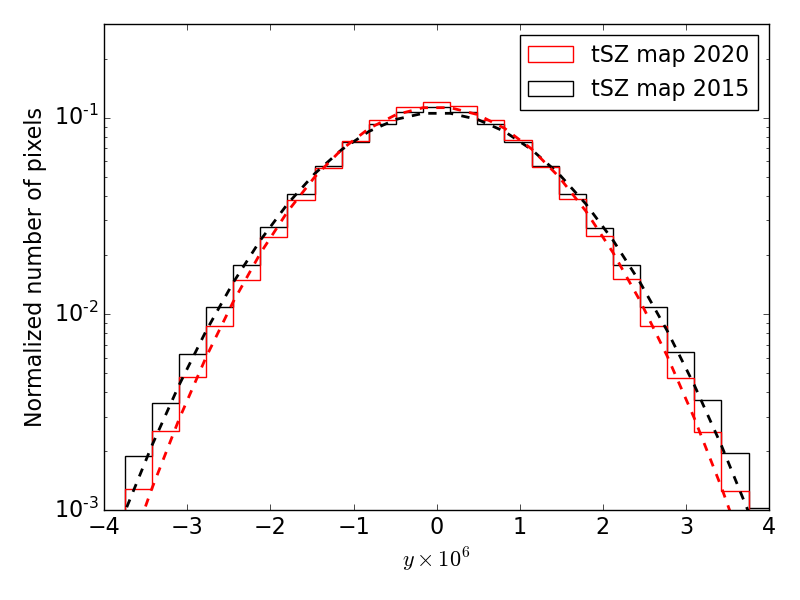}
    \includegraphics[width=0.49\linewidth]{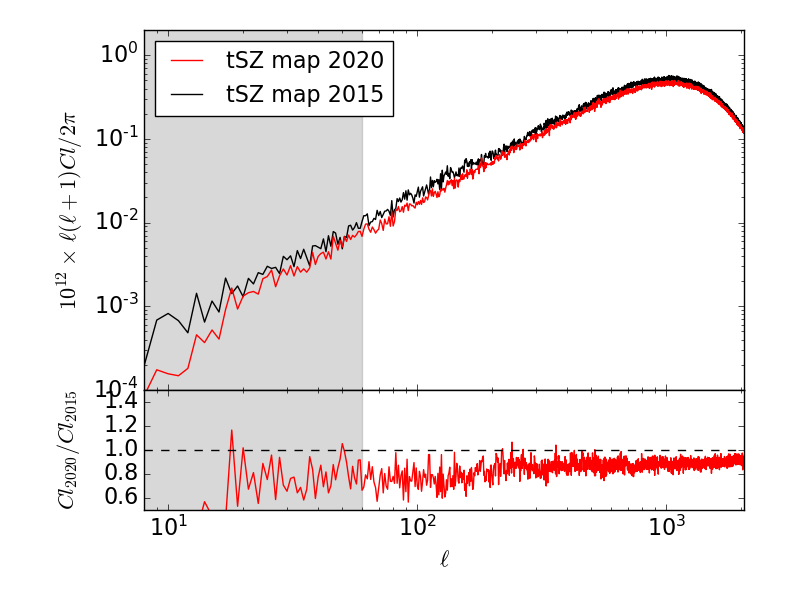}
    \caption{{\it Left}: Histograms of the 2020 and 2015 $y$-noise maps in red and black, respectively. The y-axis of the histograms is shown on a logarithmic scale. These histograms are fitted with Gaussian functions and the results are shown in dashed lines. {\it Right}: Angular power spectra of the 2020 and 2015 $y$-noise maps in red and black, respectively (see Sect. \ref{subsec:ynoise}). The gray area represents the multipole range at $\ell$ < 60, where the overall transmission is not one in the filters in Fig.~\ref{fig:scale-filter}. This multipole range is not used for our cosmological analysis. }
    \label{fig:ynoise}
    \end{figure*}

The difference of the first and last $y$-maps can be used to estimate the noise in the tSZ map (hereafter referred to as {\it $y$-noise map}). We masked the $y$-noise map with the point-source mask and 40\% galactic mask provided by \cite{Planck2016XXII}\footnote{$\sim$50\% of the sky region is available with this mask. This mask is provided from PR2 at http://pla.esac.esa.int/pla/\#home.} and compared the noise distribution of the 2020 and 2015 $y$-maps in Fig.~\ref{fig:ynoise} (left panel). They show Gaussian distributions centered at zero, and their standard deviations are $1.15 \times 10^{-6}$ for the 2020 $y$-noise map and $1.23 \times 10^{-6}$ for the 2015 $y$-noise map. It indicates that the noise is reduced by $\sim7\%$ in the new 2020 $y$-map thanks to the 8\% increased data in the original band maps.
We also checked the scale-dependent noise distribution with the angular power spectrum of the $y$-noise map in Fig.~\ref{fig:ynoise} (right panel) and found a lower noise for the 2020 $y$-map at all angular scales.

\section{tSZ Angular power spectrum}
\label{sec:anaps}

We followed \cite{Planck2014XXI} and \cite{Planck2016XXII} for the analysis of the tSZ angular power spectrum. 

We considered the cross-angular power spectrum between the $y$-maps reconstructed from the first (F) and last (L) halves of the data, named MILCA F/L (hereafter referred to as {\it 2020 cross-power spectrum}). In doing so, the bias induced by the noise in the auto-angular power spectrum can be avoided. Note that NILC F/L was used for a cosmological analysis in \citealt{Planck2014XXI} and NILC-MILCA F/L in \citealt{Planck2016XXII} (hereafter referred to as  {\it 2015 cross-power spectrum}).
The cross-power spectrum was computed using the XSPECT method \citep{Tristram2005} with the Xpol software\footnote{https://gitlab.in2p3.fr/tristram/Xpol}. The XSPECT corrects for beam convolution, pixelization, and mode-coupling induced by masks. We used the same multipole binning scheme and the same mask in \cite{Planck2016XXII}\footnote{The point-source and 50\% galactic mask provided from PR2. $\sim$42\% of the sky region is available with this mask .}. 
Uncertainties in the spectrum were also computed analytically with the Xpol software, providing the Gaussian term of uncertainties.
We additionally computed and included the trispectrum term in the uncertainties, which is the harmonic-space four-point function, to evaluate uncertainties accurately. The trispectrum term is dominant at low multipoles, as shown in \cite{Bolliet2018, Salvati2018}. 

\begin{figure}
    \centering
    \includegraphics[width=\linewidth]{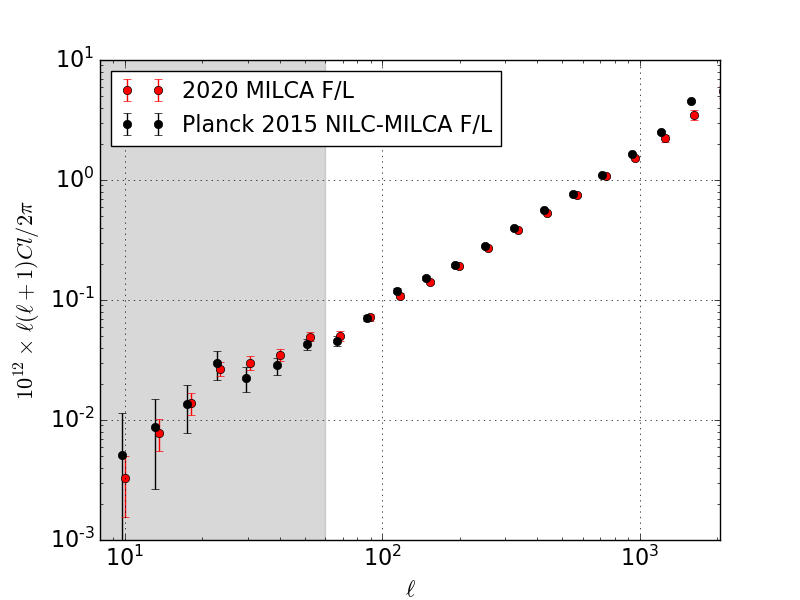}
    \caption{Angular cross-power spectra of the 2020 MILCA F/L $y$-maps and 2015 NILC-MILCA F/L $y$-maps. The sky region is masked with the 50\% Galactic and point-source mask. Only the Gaussian term of uncertainty is included here to compare with the 2015 cross-power spectrum. The 2015 NILC-MILCA F/L cross-power spectrum is shifted left along the $\ell$-axis by 3\% for a visualization purpose. The gray area represents the multipole range at $\ell$ < 60, where the overall transmission is not one in the filters in Fig.~\ref{fig:scale-filter}. This multipole range is not used for our cosmological analysis.}
    \label{fig:dlyy2015}
    \end{figure}

Figure.~\ref{fig:dlyy2015} is the comparison between the 2020 and 2015 cross-power spectra. 
The figure shows that their signal amplitudes are consistent with each other, suggesting that the tSZ signals are extracted similarly well in both $y$-maps. It also shows that the uncertainties are smaller at large scales in the 2020 cross-power spectrum than in the 2015 case. It is possibly due to the combination of several effects: lower noise, reduced survey stripes, and our new window function that suppresses large scales to reduce the residual foreground emissions, mainly from the Galactic thermal dust emission (see Appendix. \ref{sec:dust}). 

\section{Cosmological analysis}
\label{sec:cosmo}

\subsection{tSZ model}
\label{subsec:szmodel}

We modeled the tSZ power spectrum by following Section 2.2 in \cite{Salvati2018}, with a halo model \citep{Cooray2002} including the one- and two-halo terms. We used the mass function from \cite{Tinker2008} and the pressure profile model from \cite{Planck2013V} (P13). Later in Sect. \ref{subsec:systematics}, we also test two other pressure profile models : one based on the combination of \xmm\ measurements and numerical simulations (\citealt{Arnaud2010}; A10) and the other based on the analysis of the PACT data, a combination of \planck\ and ACT (\citealt{Pointecouteau2021}; PACT21).
Note that we did not vary the scaling relation between the integrated  $y$ parameter, noted $Y$, and mass as \cite{Salvati2018} did but instead fixed to the scaling relation in the pressure profile model. \cite{Arnaud2010} measured a slight deviation of 0.12 in a power-law index from the self-similar scaling relation, which is included in our model. Finally, we integrated the contribution of halos in the redshift range from 0 to 3 and the mass range from $10^{13}$ \msun to $5 \times 10^{15}$ \msun.

The tSZ power spectrum provides independent constraints on cosmological parameters. 
In particular, it is sensitive to the normalization of the matter power spectrum, commonly parameterized by $\sigma_8$, and to the total matter density, $\Omega_m$. It is also sensitive to other cosmological parameters, \eg the baryon density parameter $\Omega_b$, the Hubble constant $H_0$, and the primordial spectral index $n_s$. However, these variations are small relative to those of $\Omega_m$ and $\sigma_8$ \citep{Komatsu2002, Bolliet2018}. Thus, in our analysis, we varied only $\Omega_m$ and $\sigma_8$, and other cosmological parameters were fixed to the fiducial values in Table 1 in \cite{Planck2020VI} obtained from the \planck\ CMB measurements.

We included the mass bias, $b$, that accounts for a bias between the observationally deduced ($M_{obs}$) and true ($M_{true}$) mass of clusters expressed by $M_{obs} = (1 - b) \, M_{true}$. 
This mass bias influences cosmological analysis and degenerates with $\sigma_8$ and $\Omega_m$ parameters \citep{Planck2016XXIV}. Therefore, we used the mass bias prior, $b = 0.780 \pm 0.092$, from the Canadian Cluster Comparison Project (\citealt{Hoekstra2015}; CCCP). Later in Sect. \ref{subsec:systematics}, we also test two other mass bias priors: one from the Weighting the Giants weak lensing measurements (\citealt{Linden2014}; WtG) and the other from cosmological hydrodynamical simulations (\citealt{Biffi2016}; BIFFI).

\subsection{Foreground models}
\label{subsec:foremodel}

The measurement of the tSZ cross-power spectrum is affected by residual Galactic and extragalactic foreground emissions.

The dominant Galactic foreground emission in the \planck\ HFI maps is the Galactic thermal dust emission. Other Galactic emissions such as free-free, synchrotron, and spinning dust emissions are negligible \citep{Planck2020I}. In addition, we mitigate the Galactic contamination by restricting our cosmological analysis to $\ell > 60$. In this multipole range, we did not find significant differences in the tSZ cross-power spectrum by changing the size of the Galactic masks in Appendix.~\ref{fig:dlyydust}. Thus in the following, we do not account for the Galactic foreground emissions in our model. 

Extragalactic foreground emissions include radio and IR point sources and CIB emissions. They contribute to an excess power at small angular scales. To deal with these sources of contamination, we used physically motivated models of the foreground components.  We used \cite{Delabrouille2013} for the radio and IR point-source models and \cite{Maniyar2021} for the CIB model. We also applied the same flux limit for the masked point sources as in the \planck\ PR2 mask \citep{Planck2014XXVIII}. Based on these models, we simulated the foreground maps at the six \planck\ HFI frequencies between 100 and 857 GHz, applied the same ILC weights used for the tSZ reconstruction to the simulated foreground maps, and computed the power spectra representing the foreground residuals. The amplitude of the foreground residual power spectra was then jointly fitted with the cosmology-dependent tSZ model. In this study, we assumed the foreground model uncertainties to be 50\%, similarly to \cite{Planck2016XXII}. 

\subsection{Maximum likelihood analysis}

Cosmological constraints can be obtained by fitting the tSZ power spectrum measurement with the tSZ and foreground model.
In the measurement, we evaluated the 2020 cross-power spectrum to minimize the instrumental noise bias (see Sect. \ref{sec:anaps}).

In the model, we considered four components: tSZ, CIB, radio point sources, and infrared point sources. 
Following \cite{Planck2016XXII}, we restricted our analysis to $\ell > 60$ given that we filtered out the low multipoles to minimize the residual Galactic thermal dust contamination (see Sect. \ref{sec:mapmake}). 
We also restricted our analysis to $\ell < 1411$ because the noise is dominant at small scales. 
Finally, the observed cross-power spectrum, $C_{\ell}^{\rm obs}$, is modeled as:
\beq
C_{\ell}^{\rm obs} = C_{\ell}^{\rm tSZ} (\Omega_{m}, \sigma_{8}, b) + A_{\rm CIB} \, C_{\ell}^{\rm CIB} + A_{\rm IR} \, C_{\ell}^{\rm IR} + A_{\rm rad} \, C_{\ell}^{\rm rad} + A_{\rm CN} \, C_{\ell}^{\rm CN},
\label{eq:psmodel}
\eeq
where $C_{\ell}^{\rm tSZ} (\Omega_{m}, \sigma_{8}, b)$ is the tSZ power spectrum, $C_{\ell}^{\rm CIB}$ is the CIB power spectrum, $C_{\ell}^{\rm IR}$ and $C_{\ell}^{\rm rad}$ are the infrared and radio source power spectrum and $C_{\ell}^{\rm CN}$ is an empirical model for noise. 

We performed the likelihood analysis using the CosmoMC tool\footnote{CosmoMC released in November 2016} \citep{Lewis2002}. We used flat priors for $\Omega_m$ (0.1 to 0.5) and $\sigma_8$ (0.6 to 1.0) and Gaussian priors for $A_{\rm CIB}$, $A_{\rm IR}$, $A_{\rm rad}$, and $A_{\rm CN}$ with the center of one and standard deviation of 0.5. As a reminder, we also used the mass bias prior, $1-b = 0.780 \pm 0.092$, from the Canadian Cluster Comparison Project (\citealt{Hoekstra2015}; CCCP), including two other mass bias priors from WtG and BIFFI in Sect. \ref{subsec:systematics}. The sampling parameters and priors used in our cosmological analysis are summarized in Table.~\ref{table:cosmop}.

\begin{table}
\caption{Sampling parameters and priors.}
\begin{center}
\begin{tabular}{lll} \hline
Parameter & Symbol & Prior \\ \hline
Matter density fluctuation amplitude & $\sigma_8$ & [0.6 - 1.0] \\ 
Matter density & $\Omega_m$ & [0.1 - 0.5] \\ \hline
CIB contamination & $A_{\rm CIB}$ & $N$(1, 0.5) \\
IR-source contamination & $A_{\rm IR}$ & $N$(1, 0.5) \\
Radio-source contamination & $A_{\rm rad}$ & $N$(1, 0.5) \\
Noise & $A_{\rm CN}$ & $N$(1, 0.5) \\ 
Mass bias & $b$ & varied \\ \hline
\end{tabular}
\end{center}
\footnotesize{Sampling cosmological parameters, $\sigma_8$ and $\Omega_m$, and nuisance parameters are listed. 
Nuisance parameters include $A_{\rm CIB}$, $A_{\rm IR}$, $A_{\rm rad}$, $A_{\rm CN}$, and $b$.
[min - max] corresponds to a flat prior with minimum and maximum values.
$N$(1.0, 0.5) corresponds to a Gaussian prior with mean, 1.0, and variance, 0.5.}
\label{table:cosmop}
\end{table}

\subsection{Best-fit parameters and tSZ power spectrum}

\begin{figure}
    \centering
    \includegraphics[width=\linewidth]{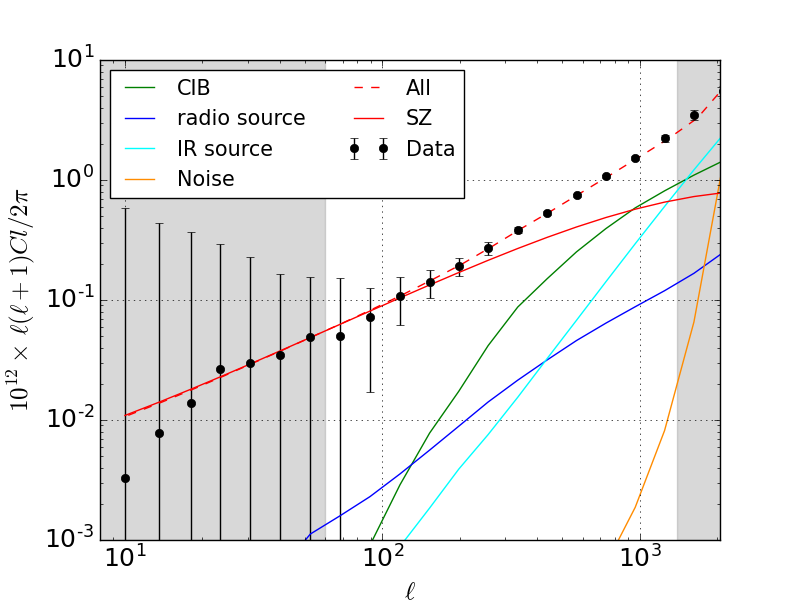}
    \caption{2020 cross-power spectrum with the 1$\sigma$ uncertainties (black) fitted with the tSZ (red) and foreground models. The fitting was performed at $60 < \ell < 1411$, with the data points outside the gray region. The considered foregrounds are CIB (green), radio sources (blue), infrared sources (cyan) and noise (orange). The sum of the tSZ and foreground models is shown in a red dashed line.}
    \label{fig:mcmcfit}
    \end{figure}

Figure~\ref{fig:mcmcfit} shows the 2020 cross-power spectrum with the uncertainties, including both the Gaussian and trispectrum terms in black. 
We fitted the measured cross-power spectrum with the tSZ model in Sect. \ref{subsec:szmodel} and foreground models in Sect. \ref{subsec:foremodel}. 
The best-fit tSZ model is presented in red. The best-fit foreground models are shown for the CIB in green, radio sources in blue, infrared sources in cyan, and noise in orange. The sum of the tSZ and foreground models is shown in a red dashed line. We found that the 2020 cross-power spectrum is dominated by tSZ at $\ell < 600$.

\begin{figure}
    \centering
    \includegraphics[width=\linewidth]{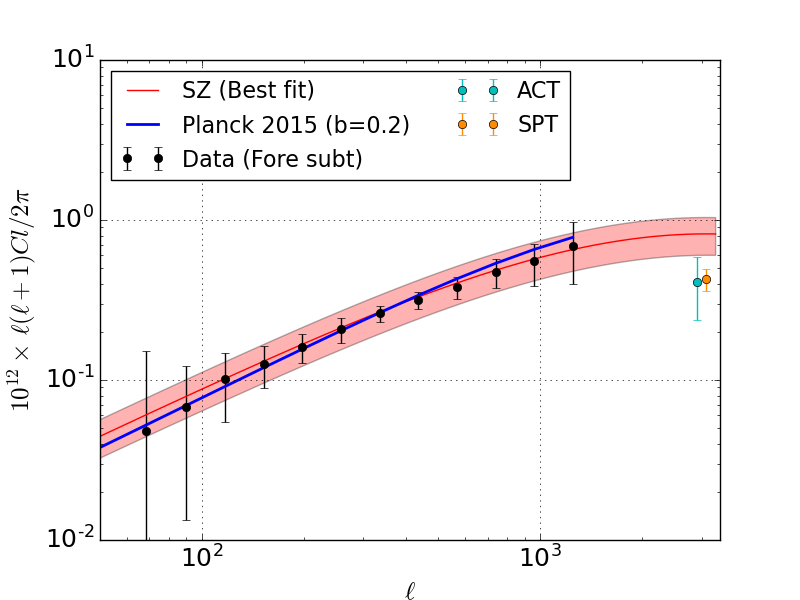}
    \caption{2020 cross-power spectrum after foreground subtraction at 60 < $\ell$ < 1411 with the 1$\sigma$ uncertainties (fore-subt:black). Our best-fit tSZ model with the 1$\sigma$ uncertainties (red) is compared to the best-fit tSZ model in \protect\cite{Planck2016XXII} for their fiducial case ($b$=0.2), the Atacama Cosmology Telescope (ACT:cyan) power spectra and the South Pole Telescope (SPT:orange) power spectra at $\ell=3000$.}
    \label{fig:mcmcbestsz}
    \end{figure}

Figure~\ref{fig:mcmcbestsz} presents the 2020 cross-power spectrum after foreground subtraction with the 1$\sigma$ uncertainties in black and our best-fit tSZ model with the 1$\sigma$ uncertainties in red (see . Table.~\ref{table:bestsz} for the actual values). 
We compared our result with the best-fit tSZ model in \cite{Planck2016XXII} for their fiducial case ($b$=0.2) and found that these results are fully consistent with each other. A slight difference is seen in the shape. It is probably due to the difference in the pressure profile models: the \planck\ analysis used \cite{Arnaud2010}, while we used \cite{Planck2013V}.
Furthermore, we show the tSZ power spectrum estimates at high multipoles by the Atacama Cosmology Telescope (ACT; cyan, \citealt{Dunkley2013}) and the South Pole Telescope (SPT; orange, \citealt{Reichardt2021}). (Note that \cite{Douspis2021} recently updated the SPT's result by including a cosmological dependence in the tSZ modeling, but we use the original SPT result in this paper.) We found that our best-fit tSZ model extrapolated to $\ell=3000$ is consistent with the ACT measurement (1.5$\sigma$ below) and SPT measurement (1.7$\sigma$ below) within 2$\sigma$ uncertainties. 

\begin{table*}
\caption{Multipole bin, measured tSZ power spectrum after foreground subtraction, its statistical uncertainties, and best-fit tSZ power spectrum.} 
\label{table:bestsz}      
\centering                          
\begin{tabular}{c c c c c c}        
\hline\hline                 
$\ell_{\rm min}$ & $\ell_{\rm max}$ & $\ell_{\rm eff}$ & $\ell(\ell+1)C_{\ell}^{yy, \rm data}/2\pi$ & $\sigma$  & $\ell(\ell+1)C_{\ell}^{yy, \rm best-fit}/2\pi$\\  
 & & & [$10^{12} y^2$] & [$10^{12} y^2$] & [$10^{12} y^2$] \\
\hline                        
60 & 78 & 68.5 & 0.048 & 0.103 & 0.063 \\  
78 & 102 & 89.5 & 0.068 & 0.055 & 0.081 \\  
102 & 133 & 117.0 & 0.101 & 0.046 & 0.104 \\  
133 & 173 & 152.5 & 0.127 & 0.038 & 0.133 \\  
173 & 224 & 198.0 & 0.161 & 0.033 & 0.169 \\  
224 & 292 & 257.5 & 0.208 & 0.037 & 0.214 \\  
292 & 380 & 335.5 & 0.261 & 0.029 & 0.269 \\  
380 & 494 & 436.5 & 0.318 & 0.039 & 0.333 \\  
494 & 642 & 567.5 & 0.380 & 0.060 & 0.406 \\  
642 & 835 & 738.0 & 0.472 & 0.097 & 0.487 \\  
835 & 1085 & 959.5 & 0.550 & 0.162 & 0.571 \\  
1085 & 1411 & 1247.5 & 0.686 & 0.285 & 0.654 \\  
\hline                                   
\end{tabular}
\end{table*}

\begin{figure}
    \centering
    \includegraphics[width=\linewidth]{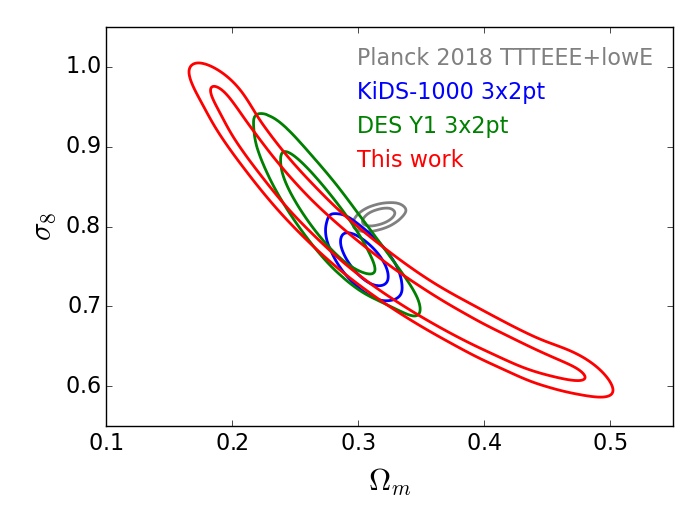}
    \caption{Posterior distribution of the cosmological parameters, $\sigma_8$ and $\Omega_m$, with 68\% and 95\% confidence interval contours obtained from our cosmological analysis. It is compared with the \planck\ CMB's (gray), KiDS-1000 3x2pt (blue), and DES Y1 3x2pt results (green). }
    \label{fig:mcmc_oms8}
    \end{figure}

Figure~\ref{fig:mcmc_oms8} shows posterior distributions of the cosmological parameters, $\sigma_8$ and $\Omega_m$, in red with 68\% and 95\% confidence interval contours. This result is compared with the one from the \planck\ CMB measurement in gray. While $\sigma_8$ and $\Omega_m$ are degenerate in our analysis, we obtained $S_8 \equiv \sigma_8 (\Omega_m / 0.3)^{0.5} = 0.764_{-0.018}^{+0.015}$. This value is lower than the \planck\ CMB's result of $S_8 = 0.830 \pm 0.013$ by $\sim 1.7 \sigma$. 

\begin{figure}
    \centering
    \includegraphics[width=\linewidth]{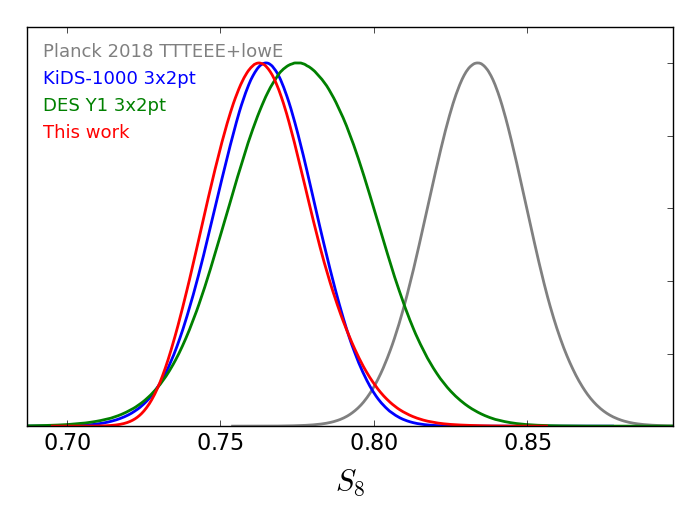}
    \caption{Posterior distribution of the $S_8$ cosmological parameter obtained from our cosmological analysis. It is compared with the \planck\ CMB's (gray), KiDS-1000 3x2pt (blue), and DES Y1 3x2pt results (green). }
    \label{fig:mcmc_s8}
    \end{figure}

We also compared our result with a joint cosmological analysis of weak gravitational lensing observations from the Kilo-Degree Survey (KiDS-1000), combined with redshift-space galaxy clustering observations from the Baryon Oscillation Spectroscopic Survey (BOSS) and galaxy-galaxy lensing observations from the overlap between KiDS-1000, BOSS, and the spectroscopic 2-degree Field Lensing Survey (2dFLenS), resulting in $S_8 = 0.766_{-0.014}^{+0.020}$ (KiDS-1000 3x2pt in blue) \citep{Heymans2021}. 
Moreover, we compared with a combined cosmological analysis of weak gravitational lensing observations from the first year of the Dark Energy Survey (DES Y1), combined with their galaxy clustering and galaxy-galaxy lensing observations, resulting in $S_8 = 0.783_{-0.025}^{+0.021}$ (DES Y1 3x2pt in green) \citep{DES2018}. 
Figure~\ref{fig:mcmc_s8} shows consistency between these weak lensing results and ours in the $S_8$ parameter, while there is a slight tension with the \planck\ CMB's result. 

Furthermore, we compared our result with the one in \cite{Planck2016XXII}. They used a different parameterization and obtained $\sigma_8 (\Omega_m / 0.28)^{3/8} = 0.80_{-0.03}^{+0.01}$ for their fiducial case ($b$=0.2). Using the same parameterization, we obtained $\sigma_8 (\Omega_m / 0.28)^{3/8} = 0.78_{-0.03}^{+0.02}$ (and also $\sigma_8 (\Omega_m / 0.28)^{3/8} = 0.793_{-0.035}^{+0.025}$ with the A10 pressure profile model used in \citealt{Planck2016XXII}) and found consistent results as expected given the full agreement in the tSZ model between \cite{Planck2016XXII} and ours in Fig~\ref{fig:mcmcbestsz}. 

Note that full posterior distributions of our parameters are shown in Fig~\ref{fig:mcmcpost}, including nuisance foreground parameters. The foreground parameters were estimated to be $A_{\rm cib}=0.82\pm0.16$, $A_{\rm ir}=1.18\pm0.34$, $A_{\rm rad}=1.02_{-0.52}^{+0.45}$, and $A_{\rm CN}=1.03\pm0.47$. 

\subsection{Systematic effects}
\label{subsec:systematics}

Finally, we consider systematic uncertainties in our cosmological analysis introduced by the mass bias and the pressure profile model. 

\begin{figure}
    \centering
    \includegraphics[width=\linewidth]{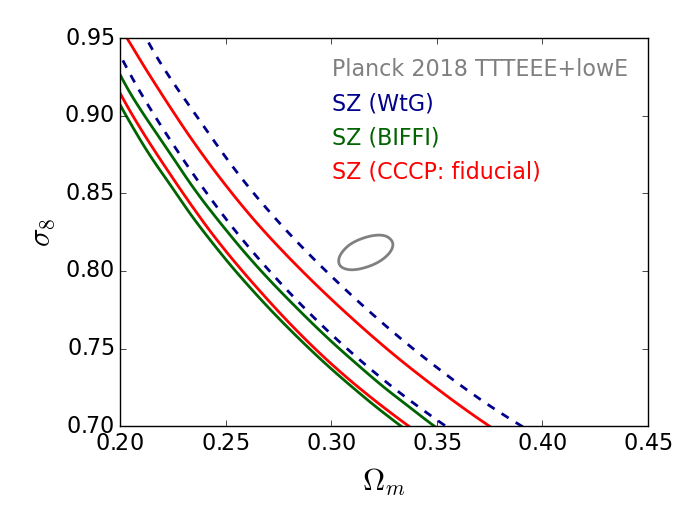}
    \caption{Posterior distribution of the cosmological parameters, $\sigma_8$ and $\Omega_m$, with 68\% confidence interval contours obtained from the \planck\ CMB analysis in gray and our tSZ analysis with three different mass bias priors (see Sect. \ref{subsec:systematics} and Table.~\ref{table:s8mbias}). Our fiducial model of CCCP is shown in red. It is compared with the WtG model in blue and BIFFI model in green. }
    \label{fig:mcmc_oms8_mbias}
    \centering
    \includegraphics[width=\linewidth]{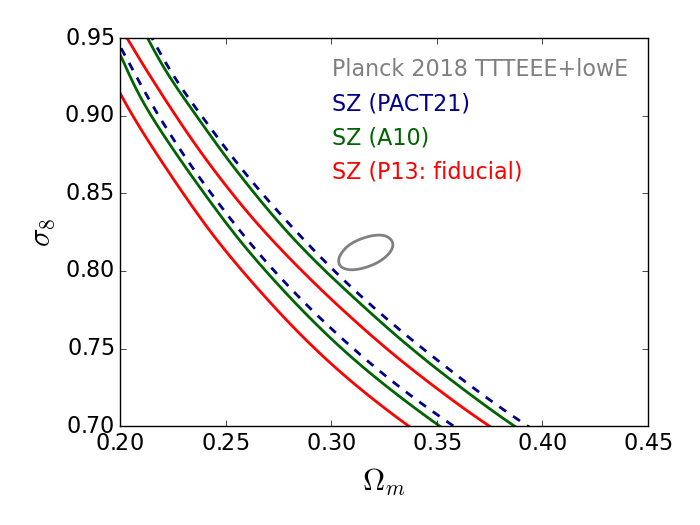}
    \caption{Posterior distribution of the cosmological parameters, $\sigma_8$ and $\Omega_m$, with 68\% confidence interval contours obtained from the \planck\ CMB analysis in gray and our tSZ analysis with three different pressure profile models (see Sect. \ref{subsec:systematics} and Table.~\ref{table:s8pressure}). Our fiducial model of PA13 is shown in red. It is compared with the PACT21 model in blue and A10 model in green.  }
    \label{fig:mcmc_oms8_pressure}
    \end{figure}

As discussed in \cite{Planck2016XXII}, the mass bias is known to affect the cosmological analysis with the tSZ power spectrum. To investigate its impact on our cosmological analysis, we replaced the CCCP's mass bias prior with two other mass bias values from WtG and BIFFI. Figure~\ref{fig:mcmc_oms8_mbias} shows their comparison in $\sigma_8$ and $\Omega_m$, and the $S_8$ values are quantitatively represented in Table.~\ref{table:s8mbias}. The result shows that the $S_8$ value increases as the mass bias increases (or $1-b$ decrease). The same trend was seen in the cosmological analysis with the \planck\ tSZ power spectrum (see Table 19 in \citealt{Planck2016XXII}) and tSZ cluster counts (see Table 7 in \citealt{Planck2016XXIV}). This systematic uncertainty may cause a $\sim 1 \sigma$ shift in the $S_8$ parameter from our fiducial value. 

In addition, the pressure profile model may affect the cosmological analysis with the tSZ power spectrum, as shown in \cite{Ruppin2019}. To investigate its impact on our cosmological analysis, we replaced the pressure profile model with two other models from A10 and PACT21. Figure~\ref{fig:mcmc_oms8_pressure} shows their comparison in $\sigma_8$ and $\Omega_m$, and the $S_8$ values are quantitatively represented in Table.~\ref{table:s8pressure}. 
The PACT21 model reduces the $S_8$ discrepancy between the \planck\ CMB and our tSZ results by $\sim 1 \sigma$ relative to our fiducial model. It may imply that the discrepancy can be further reconciled or amplified by a better understanding of baryonic physics in galaxy clusters.

\begin{table}
\caption{$S_8$ values obtained with three different mass bias priors.} 
\label{table:s8mbias}      
\centering                          
\begin{tabular}{c c c}        
\hline\hline                 
Mass bias prior & $1-b$ & $S_8 \equiv \sigma_8 (\Omega_m / 0.3)^{0.5}$ \\  
\hline                        
WtG & $0.688 \pm 0.072$ & $0.781 \pm 0.016$ \\  
CCCP (fiducial) & $0.780 \pm 0.092$ & $0.764_{-0.018}^{+0.015}$ \\  
BIFFI & $0.877 \pm 0.015$ & $0.748 \pm 0.008$ \\  
\hline                                   
\end{tabular}
\newline
\newline
\caption{$S_8$ values obtained with three different pressure profile models.} 
\label{table:s8pressure}      
\centering                          
\begin{tabular}{c c}        
\hline\hline                 
Pressure profile model & $S_8 \equiv \sigma_8 (\Omega_m / 0.3)^{0.5}$ \\  
\hline                        
P13 (fiducial) & $0.764_{-0.018}^{+0.015}$ \\  
A10 & $0.779 \pm 0.017$ \\  
PACT21 & $0.785 \pm 0.008$ \\  
\hline                                   
\end{tabular}
\end{table}

Eventually, we included these systematic uncertainties in our final results and obtained $S_8 = 0.764 \, _{-0.018}^{+0.015} \, (stat) \, _{-0.016}^{+0.031} \, (sys) $.

\section{Summary and conclusion}
\label{sec:summary}

We reconstructed a new all-sky tSZ map using the 100 to 857 GHz frequency channel maps from the \planck\ data release 4. 
For the map reconstruction, we used a component separation algorithm tailored specifically for the tSZ that we further optimized to reduce the contamination from galactic dust emission. 

We tested the resulting new $y$-map extensively in terms of noise and foreground contamination. 
Noise is reduced by $\sim$7\%, and survey stripes were minimal in this new $y$-map compared to the previous version in 2015 \citep{Planck2016XXII}. 
We confirmed that the Galactic thermal dust emission is dominant at large angular scales and radio and infrared point sources at small angular scales. Their contribution can be reduced using the Galactic and point-source mask provided in the \planck\ PR2 release. 
However, it was shown that the Galactic thermal dust emission was still not negligible in the 2015 \planck\ MILCA $y$-map even after applying a 50\% Galactic mask. Thus, we applied an extended filter at the corresponding scales to further reduce the galactic dust contamination. The CIB contamination is also not negligible at small angular scales. As a consequence of our analysis with the tSZ angular power spectrum, we found that the CIB is dominant only at $\ell > 1000$.  

We finally performed a cosmological analysis of the obtained signal. For this purpose, we constructed two $y$-maps using the first and last half-ring maps and computed their cross-power spectrum to avoid the bias induced by the noise in the auto-angular power spectrum. We used the multipole range from 60 < $\ell$ < 1411 for our cosmological analysis. The analysis of the tSZ power spectrum allowed us to set constraints on cosmological parameters, mainly of $S_8 \equiv \sigma_8 (\Omega_m / 0.3)^{0.5}$, and we obtained $S_8 = 0.764 \, _{-0.018}^{+0.015} \, (stat) \, _{-0.016}^{+0.031} \, (sys) $, in which the systematic uncertainty includes the impact of the mass bias and pressure profile model. The $S_8$ value may differ by $\pm0.16$ depending on the hydrostatic mass bias model and by $+0.21$ depending on the pressure profile model used for the analysis. Our obtained value is fully consistent with recent weak lensing results from KiDS and DES. It is also consistent with the \planck\ CMB's result \citep{Planck2020VI} within 2$\sigma$, while it is slightly lower by $\sim$1.7$\sigma$.

\section*{Acknowledgements}

{\bf The authors thank the referee for useful comments.} This research has been supported by the funding for the ByoPiC project from the European Research Council (ERC) under the European Union's Horizon 2020 research and innovation programme grant agreement ERC-2015-AdG 695561. The authors thank Guillaume Hurier for advice on the implementation of MILCA algorithm. They also Matthieu Tristram and Reijo Keskitalo for their information on the \planck\ PR4 data. The authors acknowledge fruitful discussions with the members of the ByoPiC project (https://byopic.eu/team). This publication used observations obtained with \planck\ \footnote{http://www.esa.int/Planck}, an ESA science mission with instruments and contributions directly funded by ESA Member States, NASA, and Canada. 

\section*{Data Availability Statement}

The study is based on Planck channel maps and noise maps together with the Planck sky mode simulations all available from the Planck legacy archive \url{http://pla.esac.esa.int/pla/#home}. The data produced in this study will be made available through the IDOC center \url{https://idoc.ias.universite-paris-saclay.fr/} and ByoPiC project website \url{https://byopic.eu/}.



\bibliographystyle{mnras}
\bibliography{ymap} 

\begin{thebibliography}{}
\makeatletter
\relax
\def\mn@urlcharsother{\let\do\@makeother \do\$\do\&\do\#\do\^\do\_\do\%\do\~}
\def\mn@doi{\begingroup\mn@urlcharsother \@ifnextchar [ {\mn@doi@}
  {\mn@doi@[]}}
\def\mn@doi@[#1]#2{\def\@tempa{#1}\ifx\@tempa\@empty \href
  {http://dx.doi.org/#2} {doi:#2}\else \href {http://dx.doi.org/#2} {#1}\fi
  \endgroup}
\def\mn@eprint#1#2{\mn@eprint@#1:#2::\@nil}
\def\mn@eprint@arXiv#1{\href {http://arxiv.org/abs/#1} {{\tt arXiv:#1}}}
\def\mn@eprint@dblp#1{\href {http://dblp.uni-trier.de/rec/bibtex/#1.xml}
  {dblp:#1}}
\def\mn@eprint@#1:#2:#3:#4\@nil{\def\@tempa {#1}\def\@tempb {#2}\def\@tempc
  {#3}\ifx \@tempc \@empty \let \@tempc \@tempb \let \@tempb \@tempa \fi \ifx
  \@tempb \@empty \def\@tempb {arXiv}\fi \@ifundefined
  {mn@eprint@\@tempb}{\@tempb:\@tempc}{\expandafter \expandafter \csname
  mn@eprint@\@tempb\endcsname \expandafter{\@tempc}}}

\bibitem[\protect\citeauthoryear{{Aghanim} et~al.,}{{Aghanim}
  et~al.}{2019}]{Aghanim2019}
{Aghanim} N.,  et~al., 2019, \mn@doi [\aap] {10.1051/0004-6361/201935271},
  \href {https://ui.adsabs.harvard.edu/abs/2019A&A...632A..47A} {632, A47}

\bibitem[\protect\citeauthoryear{{Arnaud}, {Pratt}, {Piffaretti},
  {B{\"o}hringer}, {Croston}  \& {Pointecouteau}}{{Arnaud}
  et~al.}{2010}]{Arnaud2010}
{Arnaud} M.,  {Pratt} G.~W.,  {Piffaretti} R.,  {B{\"o}hringer} H.,  {Croston}
  J.~H.,   {Pointecouteau} E.,  2010, \mn@doi [\aap]
  {10.1051/0004-6361/200913416}, \href
  {https://ui.adsabs.harvard.edu/abs/2010A&A...517A..92A} {517, A92}

\bibitem[\protect\citeauthoryear{{Berezhiani}, {Dolgov}  \&
  {Tkachev}}{{Berezhiani} et~al.}{2015}]{Berezhiani2015}
{Berezhiani} Z.,  {Dolgov} A.~D.,   {Tkachev} I.~I.,  2015, \mn@doi [\prd]
  {10.1103/PhysRevD.92.061303}, \href
  {https://ui.adsabs.harvard.edu/abs/2015PhRvD..92f1303B} {92, 061303}

\bibitem[\protect\citeauthoryear{{Biffi} et~al.,}{{Biffi}
  et~al.}{2016}]{Biffi2016}
{Biffi} V.,  et~al., 2016, \mn@doi [\apj] {10.3847/0004-637X/827/2/112}, \href
  {https://ui.adsabs.harvard.edu/abs/2016ApJ...827..112B} {827, 112}

\bibitem[\protect\citeauthoryear{{Bleem} et~al.,}{{Bleem}
  et~al.}{2021}]{Bleem2021}
{Bleem} L.~E.,  et~al., 2021, arXiv e-prints, \href
  {https://ui.adsabs.harvard.edu/abs/2021arXiv210205033B} {p. arXiv:2102.05033}

\bibitem[\protect\citeauthoryear{{Bolliet}, {Comis}, {Komatsu}  \&
  {Mac{\'\i}as-P{\'e}rez}}{{Bolliet} et~al.}{2018}]{Bolliet2018}
{Bolliet} B.,  {Comis} B.,  {Komatsu} E.,   {Mac{\'\i}as-P{\'e}rez} J.~F.,
  2018, \mn@doi [\mnras] {10.1093/mnras/sty823}, \href
  {https://ui.adsabs.harvard.edu/abs/2018MNRAS.477.4957B} {477, 4957}

\bibitem[\protect\citeauthoryear{{Bonjean}}{{Bonjean}}{2020}]{Bonjean2020}
{Bonjean} V.,  2020, \mn@doi [\aap] {10.1051/0004-6361/201936919}, \href
  {https://ui.adsabs.harvard.edu/abs/2020A&A...634A..81B} {634, A81}

\bibitem[\protect\citeauthoryear{{Bourdin}, {Baldi}, {Kozmanyan}  \&
  {Mazzotta}}{{Bourdin} et~al.}{2020}]{Bourdin2020}
{Bourdin} H.,  {Baldi} A.~S.,  {Kozmanyan} A.,   {Mazzotta} P.,  2020, in
  European Physical Journal Web of Conferences. p. 00007,
  \mn@doi{10.1051/epjconf/202022800007}

\bibitem[\protect\citeauthoryear{{Chiang}, {Makiya}, {M{\'e}nard}  \&
  {Komatsu}}{{Chiang} et~al.}{2020}]{Chiang2020}
{Chiang} Y.-K.,  {Makiya} R.,  {M{\'e}nard} B.,   {Komatsu} E.,  2020, \mn@doi
  [\apj] {10.3847/1538-4357/abb403}, \href
  {https://ui.adsabs.harvard.edu/abs/2020ApJ...902...56C} {902, 56}

\bibitem[\protect\citeauthoryear{{Cooray} \& {Sheth}}{{Cooray} \&
  {Sheth}}{2002}]{Cooray2002}
{Cooray} A.,  {Sheth} R.,  2002, \mn@doi [\physrep]
  {10.1016/S0370-1573(02)00276-4}, \href
  {https://ui.adsabs.harvard.edu/abs/2002PhR...372....1C} {372, 1}

\bibitem[\protect\citeauthoryear{{DES Collaboration}}{{DES
  Collaboration}}{2021}]{DES2021}
{DES Collaboration} 2021, arXiv e-prints, \href
  {https://ui.adsabs.harvard.edu/abs/2021arXiv210513549D} {p. arXiv:2105.13549}

\bibitem[\protect\citeauthoryear{{Dark Energy Survey Collaboration}}{{Dark
  Energy Survey Collaboration}}{2018}]{DES2018}
{Dark Energy Survey Collaboration} 2018, \mn@doi [\prd]
  {10.1103/PhysRevD.98.043526}, \href
  {https://ui.adsabs.harvard.edu/abs/2018PhRvD..98d3526A} {98, 043526}

\bibitem[\protect\citeauthoryear{{Delabrouille} et~al.,}{{Delabrouille}
  et~al.}{2013}]{Delabrouille2013}
{Delabrouille} J.,  et~al., 2013, \mn@doi [\aap] {10.1051/0004-6361/201220019},
  \href {https://ui.adsabs.harvard.edu/abs/2013A&A...553A..96D} {553, A96}

\bibitem[\protect\citeauthoryear{{Di Valentino}, {Melchiorri}, {Mena}  \&
  {Vagnozzi}}{{Di Valentino} et~al.}{2020}]{Valentino2020}
{Di Valentino} E.,  {Melchiorri} A.,  {Mena} O.,   {Vagnozzi} S.,  2020,
  \mn@doi [Physics of the Dark Universe] {10.1016/j.dark.2020.100666}, \href
  {https://ui.adsabs.harvard.edu/abs/2020PDU....3000666D} {30, 100666}

\bibitem[\protect\citeauthoryear{{Douspis}, {Salvati}, {Gorce}  \&
  {Aghanim}}{{Douspis} et~al.}{2021}]{Douspis2021}
{Douspis} M.,  {Salvati} L.,  {Gorce} A.,   {Aghanim} N.,  2021, arXiv
  e-prints, \href {https://ui.adsabs.harvard.edu/abs/2021arXiv210903272D} {p.
  arXiv:2109.03272}

\bibitem[\protect\citeauthoryear{{Dunkley} et~al.,}{{Dunkley}
  et~al.}{2013}]{Dunkley2013}
{Dunkley} J.,  et~al., 2013, \mn@doi [\jcap] {10.1088/1475-7516/2013/07/025},
  \href {https://ui.adsabs.harvard.edu/abs/2013JCAP...07..025D} {2013, 025}

\bibitem[\protect\citeauthoryear{{G{\'o}rski}, {Hivon}, {Banday}, {Wandelt},
  {Hansen}, {Reinecke}  \& {Bartelmann}}{{G{\'o}rski}
  et~al.}{2005}]{Gorski2005}
{G{\'o}rski} K.~M.,  {Hivon} E.,  {Banday} A.~J.,  {Wandelt} B.~D.,  {Hansen}
  F.~K.,  {Reinecke} M.,   {Bartelmann} M.,  2005, \mn@doi [\apj]
  {10.1086/427976}, \href
  {https://ui.adsabs.harvard.edu/abs/2005ApJ...622..759G} {622, 759}

\bibitem[\protect\citeauthoryear{{Greco}, {Hill}, {Spergel}  \&
  {Battaglia}}{{Greco} et~al.}{2015}]{Greco2015}
{Greco} J.~P.,  {Hill} J.~C.,  {Spergel} D.~N.,   {Battaglia} N.,  2015,
  \mn@doi [\apj] {10.1088/0004-637X/808/2/151}, \href
  {https://ui.adsabs.harvard.edu/abs/2015ApJ...808..151G} {808, 151}

\bibitem[\protect\citeauthoryear{{Haehnelt} \& {Tegmark}}{{Haehnelt} \&
  {Tegmark}}{1996}]{Haehnelt1996}
{Haehnelt} M.~G.,  {Tegmark} M.,  1996, \mn@doi [\mnras]
  {10.1093/mnras/279.2.545}, \href
  {https://ui.adsabs.harvard.edu/abs/1996MNRAS.279..545H} {279, 545}

\bibitem[\protect\citeauthoryear{{Heymans} et~al.,}{{Heymans}
  et~al.}{2021}]{Heymans2021}
{Heymans} C.,  et~al., 2021, \mn@doi [\aap] {10.1051/0004-6361/202039063},
  \href {https://ui.adsabs.harvard.edu/abs/2021A&A...646A.140H} {646, A140}

\bibitem[\protect\citeauthoryear{{Hill}, {Baxter}, {Lidz}, {Greco}  \&
  {Jain}}{{Hill} et~al.}{2018}]{Hill2018}
{Hill} J.~C.,  {Baxter} E.~J.,  {Lidz} A.,  {Greco} J.~P.,   {Jain} B.,  2018,
  \mn@doi [\prd] {10.1103/PhysRevD.97.083501}, \href
  {https://ui.adsabs.harvard.edu/abs/2018PhRvD..97h3501H} {97, 083501}

\bibitem[\protect\citeauthoryear{{Hoekstra}, {Herbonnet}, {Muzzin}, {Babul},
  {Mahdavi}, {Viola}  \& {Cacciato}}{{Hoekstra} et~al.}{2015}]{Hoekstra2015}
{Hoekstra} H.,  {Herbonnet} R.,  {Muzzin} A.,  {Babul} A.,  {Mahdavi} A.,
  {Viola} M.,   {Cacciato} M.,  2015, \mn@doi [\mnras] {10.1093/mnras/stv275},
  \href {https://ui.adsabs.harvard.edu/abs/2015MNRAS.449..685H} {449, 685}

\bibitem[\protect\citeauthoryear{{Hojjati} et~al.,}{{Hojjati}
  et~al.}{2017}]{Hojjati2017}
{Hojjati} A.,  et~al., 2017, \mn@doi [\mnras] {10.1093/mnras/stx1659}, \href
  {https://ui.adsabs.harvard.edu/abs/2017MNRAS.471.1565H} {471, 1565}

\bibitem[\protect\citeauthoryear{{Hurier} \& {Lacasa}}{{Hurier} \&
  {Lacasa}}{2017}]{Hurier2017}
{Hurier} G.,  {Lacasa} F.,  2017, \mn@doi [\aap] {10.1051/0004-6361/201630041},
  \href {https://ui.adsabs.harvard.edu/abs/2017A&A...604A..71H} {604, A71}

\bibitem[\protect\citeauthoryear{{Hurier}, {Mac{\'\i}as-P{\'e}rez}  \&
  {Hildebrandt}}{{Hurier} et~al.}{2013}]{Hurier2013}
{Hurier} G.,  {Mac{\'\i}as-P{\'e}rez} J.~F.,   {Hildebrandt} S.,  2013, \mn@doi
  [\aap] {10.1051/0004-6361/201321891}, \href
  {https://ui.adsabs.harvard.edu/abs/2013A&A...558A.118H} {558, A118}

\bibitem[\protect\citeauthoryear{{Komatsu} \& {Seljak}}{{Komatsu} \&
  {Seljak}}{2002}]{Komatsu2002}
{Komatsu} E.,  {Seljak} U.,  2002, \mn@doi [\mnras]
  {10.1046/j.1365-8711.2002.05889.x}, \href
  {https://ui.adsabs.harvard.edu/abs/2002MNRAS.336.1256K} {336, 1256}

\bibitem[\protect\citeauthoryear{{Koukoufilippas}, {Alonso}, {Bilicki}  \&
  {Peacock}}{{Koukoufilippas} et~al.}{2020}]{Koukoufilippas2020}
{Koukoufilippas} N.,  {Alonso} D.,  {Bilicki} M.,   {Peacock} J.~A.,  2020,
  \mn@doi [\mnras] {10.1093/mnras/stz3351}, \href
  {https://ui.adsabs.harvard.edu/abs/2020MNRAS.491.5464K} {491, 5464}

\bibitem[\protect\citeauthoryear{{Lewis} \& {Bridle}}{{Lewis} \&
  {Bridle}}{2002}]{Lewis2002}
{Lewis} A.,  {Bridle} S.,  2002, \mn@doi [\prd] {10.1103/PhysRevD.66.103511},
  \href {https://ui.adsabs.harvard.edu/abs/2002PhRvD..66j3511L} {66, 103511}

\bibitem[\protect\citeauthoryear{{Madhavacheril} et~al.,}{{Madhavacheril}
  et~al.}{2020}]{Madhavacheril2020}
{Madhavacheril} M.~S.,  et~al., 2020, \mn@doi [\prd]
  {10.1103/PhysRevD.102.023534}, \href
  {https://ui.adsabs.harvard.edu/abs/2020PhRvD.102b3534M} {102, 023534}

\bibitem[\protect\citeauthoryear{{Makiya}, {Ando}  \& {Komatsu}}{{Makiya}
  et~al.}{2018}]{Makiya2018}
{Makiya} R.,  {Ando} S.,   {Komatsu} E.,  2018, \mn@doi [\mnras]
  {10.1093/mnras/sty2031}, \href
  {https://ui.adsabs.harvard.edu/abs/2018MNRAS.480.3928M} {480, 3928}

\bibitem[\protect\citeauthoryear{{Maniyar}, {B{\'e}thermin}  \&
  {Lagache}}{{Maniyar} et~al.}{2021}]{Maniyar2021}
{Maniyar} A.,  {B{\'e}thermin} M.,   {Lagache} G.,  2021, \mn@doi [\aap]
  {10.1051/0004-6361/202038790}, \href
  {https://ui.adsabs.harvard.edu/abs/2021A&A...645A..40M} {645, A40}

\bibitem[\protect\citeauthoryear{{Osato}, {Shirasaki}, {Miyatake}, {Nagai},
  {Yoshida}, {Oguri}  \& {Takahashi}}{{Osato} et~al.}{2020}]{Osato2020}
{Osato} K.,  {Shirasaki} M.,  {Miyatake} H.,  {Nagai} D.,  {Yoshida} N.,
  {Oguri} M.,   {Takahashi} R.,  2020, \mn@doi [\mnras]
  {10.1093/mnras/staa117}, \href
  {https://ui.adsabs.harvard.edu/abs/2020MNRAS.492.4780O} {492, 4780}

\bibitem[\protect\citeauthoryear{{Pandey}, {Karwal}  \& {Das}}{{Pandey}
  et~al.}{2020}]{Pandey2020}
{Pandey} K.~L.,  {Karwal} T.,   {Das} S.,  2020, \mn@doi [\jcap]
  {10.1088/1475-7516/2020/07/026}, \href
  {https://ui.adsabs.harvard.edu/abs/2020JCAP...07..026P} {2020, 026}

\bibitem[\protect\citeauthoryear{{Planck Collaboration}}{{Planck
  Collaboration}}{2013}]{Planck2013V}
{Planck Collaboration} 2013, \mn@doi [\aap] {10.1051/0004-6361/201220040},
  \href {https://ui.adsabs.harvard.edu/abs/2013A&A...550A.131P} {550, A131}

\bibitem[\protect\citeauthoryear{{Planck Collaboration}}{{Planck
  Collaboration}}{2014a}]{Planck2014XX}
{Planck Collaboration} 2014a, \mn@doi [\aap] {10.1051/0004-6361/201321521},
  \href {https://ui.adsabs.harvard.edu/abs/2014A&A...571A..20P} {571, A20}

\bibitem[\protect\citeauthoryear{{Planck Collaboration}}{{Planck
  Collaboration}}{2014b}]{Planck2014XXI}
{Planck Collaboration} 2014b, \mn@doi [\aap] {10.1051/0004-6361/201321522},
  \href {https://ui.adsabs.harvard.edu/abs/2014A&A...571A..21P} {571, A21}

\bibitem[\protect\citeauthoryear{{Planck Collaboration}}{{Planck
  Collaboration}}{2014c}]{Planck2014XXVIII}
{Planck Collaboration} 2014c, \mn@doi [\aap] {10.1051/0004-6361/201321524},
  \href {https://ui.adsabs.harvard.edu/abs/2014A&A...571A..28P} {571, A28}

\bibitem[\protect\citeauthoryear{{Planck Collaboration}}{{Planck
  Collaboration}}{2016a}]{Planck2016XXII}
{Planck Collaboration} 2016a, \mn@doi [\aap] {10.1051/0004-6361/201525826},
  \href {https://ui.adsabs.harvard.edu/abs/2016A&A...594A..22P} {594, A22}

\bibitem[\protect\citeauthoryear{{Planck Collaboration}}{{Planck
  Collaboration}}{2016b}]{Planck2016XXIV}
{Planck Collaboration} 2016b, \mn@doi [\aap] {10.1051/0004-6361/201525833},
  \href {https://ui.adsabs.harvard.edu/abs/2016A&A...594A..24P} {594, A24}

\bibitem[\protect\citeauthoryear{{Planck Collaboration}}{{Planck
  Collaboration}}{2016c}]{Planck2016XXVII}
{Planck Collaboration} 2016c, \mn@doi [\aap] {10.1051/0004-6361/201525823},
  \href {https://ui.adsabs.harvard.edu/abs/2016A&A...594A..27P} {594, A27}

\bibitem[\protect\citeauthoryear{{Planck Collaboration}}{{Planck
  Collaboration}}{2020a}]{Planck2020I}
{Planck Collaboration} 2020a, \mn@doi [\aap] {10.1051/0004-6361/201833880},
  \href {https://ui.adsabs.harvard.edu/abs/2020A&A...641A...1P} {641, A1}

\bibitem[\protect\citeauthoryear{{Planck Collaboration}}{{Planck
  Collaboration}}{2020b}]{Planck2020VI}
{Planck Collaboration} 2020b, \mn@doi [\aap] {10.1051/0004-6361/201833910},
  \href {https://ui.adsabs.harvard.edu/abs/2020A&A...641A...6P} {641, A6}

\bibitem[\protect\citeauthoryear{{Planck Collaboration}}{{Planck
  Collaboration}}{2020c}]{Planck2020LVII}
{Planck Collaboration} 2020c, \mn@doi [\aap] {10.1051/0004-6361/202038073},
  \href {https://ui.adsabs.harvard.edu/abs/2020A&A...643A..42P} {643, A42}

\bibitem[\protect\citeauthoryear{{Pointecouteau} et~al.,}{{Pointecouteau}
  et~al.}{2021}]{Pointecouteau2021}
{Pointecouteau} E.,  et~al., 2021, arXiv e-prints, \href
  {https://ui.adsabs.harvard.edu/abs/2021arXiv210505607P} {p. arXiv:2105.05607}

\bibitem[\protect\citeauthoryear{{Reichardt} et~al.,}{{Reichardt}
  et~al.}{2021}]{Reichardt2021}
{Reichardt} C.~L.,  et~al., 2021, \mn@doi [\apj] {10.3847/1538-4357/abd407},
  \href {https://ui.adsabs.harvard.edu/abs/2021ApJ...908..199R} {908, 199}

\bibitem[\protect\citeauthoryear{{Remazeilles}, {Delabrouille}  \&
  {Cardoso}}{{Remazeilles} et~al.}{2011}]{Remazeilles2011}
{Remazeilles} M.,  {Delabrouille} J.,   {Cardoso} J.-F.,  2011, \mn@doi
  [\mnras] {10.1111/j.1365-2966.2010.17624.x}, \href
  {https://ui.adsabs.harvard.edu/abs/2011MNRAS.410.2481R} {410, 2481}

\bibitem[\protect\citeauthoryear{{Ruppin}, {Mayet}, {Mac{\'\i}as-P{\'e}rez}  \&
  {Perotto}}{{Ruppin} et~al.}{2019}]{Ruppin2019}
{Ruppin} F.,  {Mayet} F.,  {Mac{\'\i}as-P{\'e}rez} J.~F.,   {Perotto} L.,
  2019, \mn@doi [\mnras] {10.1093/mnras/stz2669}, \href
  {https://ui.adsabs.harvard.edu/abs/2019MNRAS.490..784R} {490, 784}

\bibitem[\protect\citeauthoryear{{Salvati}, {Douspis}  \& {Aghanim}}{{Salvati}
  et~al.}{2018}]{Salvati2018}
{Salvati} L.,  {Douspis} M.,   {Aghanim} N.,  2018, \mn@doi [\aap]
  {10.1051/0004-6361/201731990}, \href
  {https://ui.adsabs.harvard.edu/abs/2018A&A...614A..13S} {614, A13}

\bibitem[\protect\citeauthoryear{{Sunyaev} \& {Zeldovich}}{{Sunyaev} \&
  {Zeldovich}}{1972}]{Sunyaev1972}
{Sunyaev} R.~A.,  {Zeldovich} Y.~B.,  1972, Comments on Astrophysics and Space
  Physics, \href {http://adsabs.harvard.edu/abs/1972CoASP...4..173S} {4, 173}

\bibitem[\protect\citeauthoryear{{Tanimura} et~al.,}{{Tanimura}
  et~al.}{2019a}]{Tanimura2019a}
{Tanimura} H.,  et~al., 2019a, \mn@doi [\mnras] {10.1093/mnras/sty3118}, \href
  {https://ui.adsabs.harvard.edu/abs/2019MNRAS.483..223T} {483, 223}

\bibitem[\protect\citeauthoryear{{Tanimura}, {Aghanim}, {Douspis}, {Beelen}  \&
  {Bonjean}}{{Tanimura} et~al.}{2019b}]{Tanimura2019b}
{Tanimura} H.,  {Aghanim} N.,  {Douspis} M.,  {Beelen} A.,   {Bonjean} V.,
  2019b, \mn@doi [\aap] {10.1051/0004-6361/201833413}, \href
  {https://ui.adsabs.harvard.edu/abs/2019A&A...625A..67T} {625, A67}

\bibitem[\protect\citeauthoryear{{Tanimura} et~al.,}{{Tanimura}
  et~al.}{2020a}]{Tanimura2020a}
{Tanimura} H.,  et~al., 2020a, \mn@doi [\mnras] {10.1093/mnras/stz3130}, \href
  {https://ui.adsabs.harvard.edu/abs/2020MNRAS.491.2318T} {491, 2318}

\bibitem[\protect\citeauthoryear{{Tanimura}, {Aghanim}, {Bonjean}, {Malavasi}
  \& {Douspis}}{{Tanimura} et~al.}{2020b}]{Tanimura2020b}
{Tanimura} H.,  {Aghanim} N.,  {Bonjean} V.,  {Malavasi} N.,   {Douspis} M.,
  2020b, \mn@doi [\aap] {10.1051/0004-6361/201937158}, \href
  {https://ui.adsabs.harvard.edu/abs/2020A&A...637A..41T} {637, A41}

\bibitem[\protect\citeauthoryear{{Tanimura}, {Aghanim}, {Kolodzig}, {Douspis}
  \& {Malavasi}}{{Tanimura} et~al.}{2020c}]{Tanimura2020c}
{Tanimura} H.,  {Aghanim} N.,  {Kolodzig} A.,  {Douspis} M.,   {Malavasi} N.,
  2020c, \mn@doi [\aap] {10.1051/0004-6361/202038521}, \href
  {https://ui.adsabs.harvard.edu/abs/2020A&A...643L...2T} {643, L2}

\bibitem[\protect\citeauthoryear{{Tinker}, {Kravtsov}, {Klypin}, {Abazajian},
  {Warren}, {Yepes}, {Gottl{\"o}ber}  \& {Holz}}{{Tinker}
  et~al.}{2008}]{Tinker2008}
{Tinker} J.,  {Kravtsov} A.~V.,  {Klypin} A.,  {Abazajian} K.,  {Warren} M.,
  {Yepes} G.,  {Gottl{\"o}ber} S.,   {Holz} D.~E.,  2008, \mn@doi [\apj]
  {10.1086/591439}, \href
  {https://ui.adsabs.harvard.edu/abs/2008ApJ...688..709T} {688, 709}

\bibitem[\protect\citeauthoryear{{Tristram}, {Mac{\'\i}as-P{\'e}rez}, {Renault}
   \& {Santos}}{{Tristram} et~al.}{2005}]{Tristram2005}
{Tristram} M.,  {Mac{\'\i}as-P{\'e}rez} J.~F.,  {Renault} C.,   {Santos} D.,
  2005, \mn@doi [\mnras] {10.1111/j.1365-2966.2005.08760.x}, \href
  {https://ui.adsabs.harvard.edu/abs/2005MNRAS.358..833T} {358, 833}

\bibitem[\protect\citeauthoryear{{de Graaff}, {Cai}, {Heymans}  \&
  {Peacock}}{{de Graaff} et~al.}{2019}]{Graaff2019}
{de Graaff} A.,  {Cai} Y.-C.,  {Heymans} C.,   {Peacock} J.~A.,  2019, \mn@doi
  [\aap] {10.1051/0004-6361/201935159}, \href
  {https://ui.adsabs.harvard.edu/abs/2019A&A...624A..48D} {624, A48}

\bibitem[\protect\citeauthoryear{{von der Linden} et~al.,}{{von der Linden}
  et~al.}{2014}]{Linden2014}
{von der Linden} A.,  et~al., 2014, \mn@doi [\mnras] {10.1093/mnras/stu1423},
  \href {https://ui.adsabs.harvard.edu/abs/2014MNRAS.443.1973V} {443, 1973}

\makeatother
\end{thebibliography}




\appendix

\section{Galactic dust contamination}
\label{sec:dust}
\begin{figure}
    \centering
    \includegraphics[width=\linewidth]{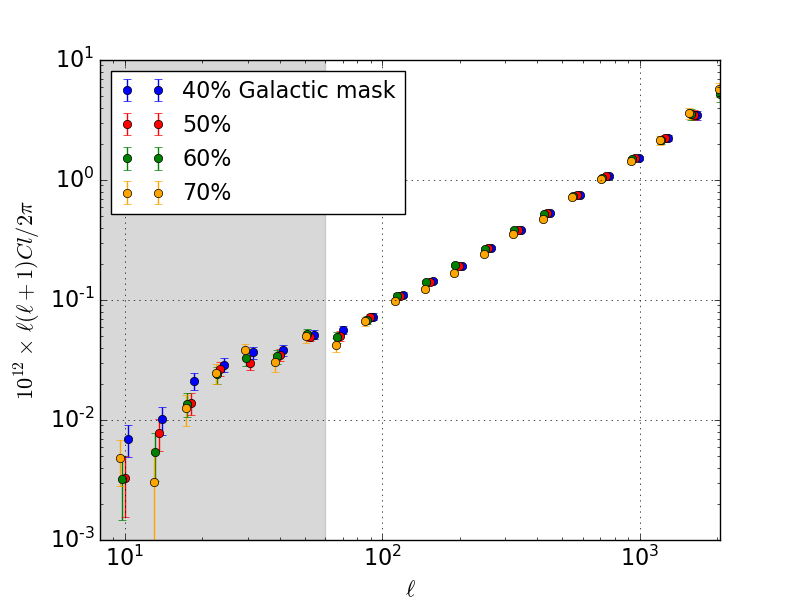}
    \caption{Angular cross-power spectra of the 2020 $y$-maps with different Galactic masks corresponding to 40\% (blue), 50\% (red), 60\% (green) and 70\% (orange) of the sky. Point sources are also masked. Only the Gaussian term of uncertainty is included here to clarify the difference. The cross-power spectra are shifted along the $\ell$-axis right by 3\% (40\%), left by 3\% (60\%), and left by 4\% (70\%) for a visualization purpose.}
    \label{fig:dlyydust}
    \end{figure}

We tested the effect of the residual Galactic thermal dust emission on the estimated signal by applying several Galactic masks provided in \cite{Planck2016XXII}: 40\%, 50\%, 60\%, and 70\% Galactic masks. The resulting Fig.~\ref{fig:dlyydust} shows that the signals at $\ell$ < 20 decrease when imposing more severe masks, implying lower contamination from the Galactic dust emission. A similar trend is seen in the cross-spectra of 2015 NILC F/L and 2015 MILCA F/L in Fig. 11 of \cite{Planck2016XXII}. That figure shows more significant dust contamination in the 2015 MILCA F/L cross-power spectrum at $\ell$ < 30, while the contamination is not significant in our $y$-map. The difference is due mainly to the optimized window function used for the reconstruction of the $y$-map. 

\section{Posterior distributions}
\label{sec:mcmcpost}

\begin{figure*}
    \centering
    \includegraphics[width=\linewidth]{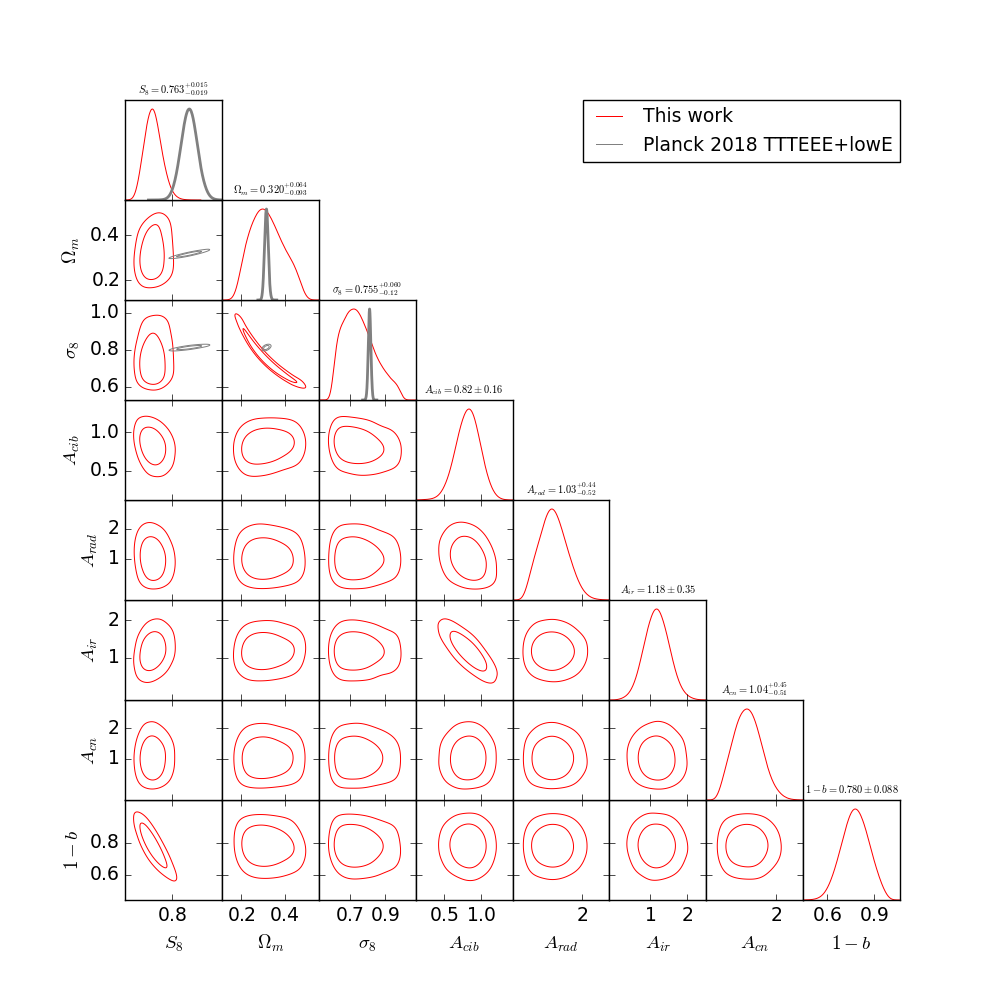}
    \caption{Posterior distribution of the cosmological parameter, $S_8$ $\equiv$ $\sigma_8 (\Omega_m / 0.3)^{0.5}$, $\sigma_8$, $\Omega_m$, the mass bias in $1-b$, and foreground parameters $A_{\rm CIB}$, $A_{\rm IR}$ and $A_{\rm rad}$ with 68\% and 95\% confidence interval contours obtained from our cosmological analysis (red). It is compared with the \planck\ CMB's result (gray). }
    \label{fig:mcmcpost}
    \end{figure*}

Fig.~\ref{fig:mcmcpost} shows one-dimensional and two-dimensional probability distributions of parameters derived from our cosmological analysis in Sect. \ref{sec:cosmo}. The best-fit values and 1$\sigma$ uncertainties are shown on top of the one-dimensional probability distributions. 

\section{7.5' $y$-map}
\label{sec:ymap7p5}

\begin{figure*}
    \centering
    \includegraphics[width=\linewidth]{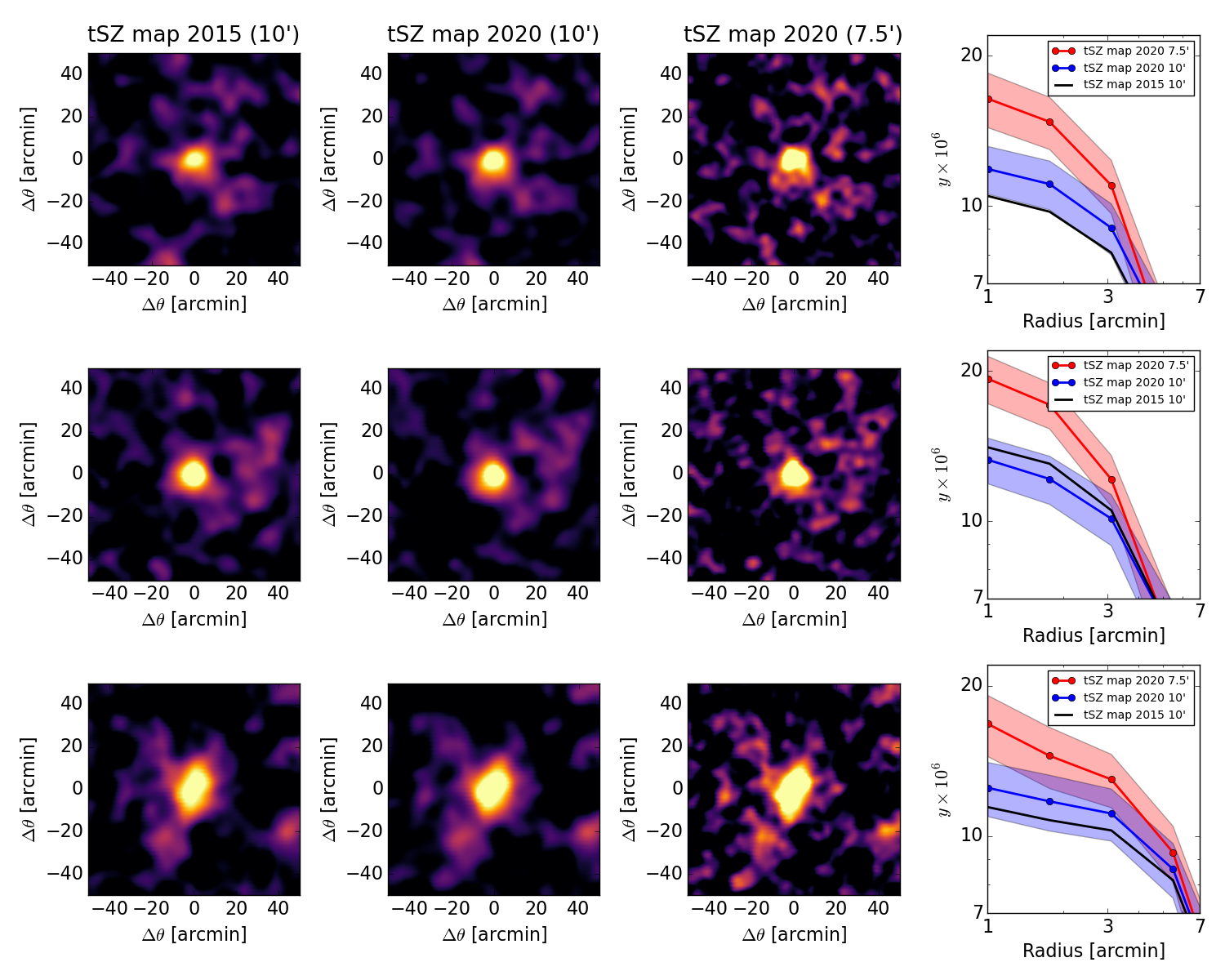}
    \caption{Two-dimensional $y$-maps centered at the position of three PSZ2 clusters corresponding to Abell 2397 at $z=0.22$ (Top row), RXC J2104.3-4120 at $z=0.16$ (middle row) and Abell 2593 $z=0.04$ (bottom row) extracted from 2015 10' (first column from left), 2020 10' (second column from left), and 2020 7.5' (third column from left) $y$-maps. Radial profiles of these clusters are computed with 2015 10' (black line), 2020 10' (blue line), and 2020 7.5' (red lines) $y$-maps and are shown on a logarithmic scale (right column). The 1$\sigma$ uncertainties of the radial profiles from the 2020 10' and 2020 7.5' $y$-maps are shown in blue and red shaded areas.}
    \label{fig:yprof7}
    \end{figure*}

We produced a $y$-map with a slightly higher angular resolution of 7.5' in the same way as described in Sect. \ref{sec:mapmake}. 
As in Fig.~\ref{fig:yprof}, we compared in Fig.~\ref{fig:yprof7} two-dimensional $y$-maps and radial $y$-profiles centered at the position of three PSZ2 clusters corresponding to Abell 2397 at $z=0.22$ (top row), RXC J2104.3-4120 at $z=0.16$ (middle row), and Abell 2593 $z=0.04$ (bottom row), using 2015 10' (first column from left), 2020 10' (second column from left) and 2020 7.5' (third column from left) $y$-maps. In the radial y-profiles, the results are consistent at large angular scales, but improvements are visible at small angular scales with the 2020 7.5' $y$-map. The 1$\sigma$ uncertainties of the radial profiles from the 2020 10' and 2020 7.5' $y$-maps are shown in blue and red shaded areas.

\bsp	
\label{lastpage}
\end{document}